\documentclass[amsmath,amssymb]{revtex4-1}

\usepackage{graphicx}
\usepackage{dcolumn}
\usepackage{bm}
\usepackage[utf8]{inputenc}
\usepackage[T1]{fontenc}
\usepackage{mathptmx}
\usepackage{etoolbox}

\usepackage{xcolor}

\makeatletter
\def\@email#1#2{%
 \endgroup
 \patchcmd{\titleblock@produce}
  {\frontmatter@RRAPformat}
  {\frontmatter@RRAPformat{\produce@RRAP{*#1\href{mailto:#2}{#2}}}\frontmatter@RRAPformat}
  {}{}
}%
\makeatother

\usepackage{enumitem}
\usepackage{bigints}
\usepackage{pifont}
\usepackage{relsize}

\usepackage[caption=false]{subfig}
\usepackage{ragged2e} 
\DeclareCaptionJustification{justified}{\justifying}

\newcommand{\Rm}{\mathcal{R}}

\newcommand{\du}{\dot{u}}

\newcommand{\pu}{\frac{\partial u }{\partial t}}
\newcommand{\ptu}{\frac{\partial \tilde{u}}{\partial s}}
\newcommand{\tu}{{\tilde{u}}}
\newcommand{\tp}{{\tilde{p}}}
\newcommand{\mC}{{\mathcal{C}}}
\newcommand{\mE}{{\mathcal{E}}}
\newcommand{\mB}{{\mathcal{B}}}
\newcommand{\mF}{{\mathcal{F}}}
\newcommand{\mP}{{\mathcal{P}}}
\newcommand{\mL}{{\mathcal{L}}}
\newcommand{\mR}{{\mathcal{R}}}
\newcommand{\lm}{(\lambda + 2 \mu)}
\newcommand{\tsigma}{\tilde{\sigma}}
\newcommand{\ttau}{\tilde{\tau}}
\newcommand{\tmF}{\tilde{\mathcal{F}}}

\newcommand{\tptree}{\tp_{\rm{tree}}\left(\tmF\left[\frac{\partial \tu}{\partial s}\right]\right)}

\newcommand{\bfer}{\mathbf{e_r}}
\newcommand{\bfeth}{\mathbf{e_{\theta}}}
\newcommand{\F}{\mathcal{F}[\dot{u}]}
\newcommand{\tF}{\tilde{\mathcal{F}}[\frac{\partial \tu}{\partial s}]}

\newcommand{\tx}{\tilde{x}}

\newcommand{\cred}{\color{black}}
\newcommand{\cb}{\color{black}}

\begin{document}

\title{Optimal efficiency of high frequency chest wall oscillations and links with resistance and compliance in a model of the lung.}

\author{Micha\"el Brunengo}
\address{Université Côte d'Azur, LJAD, Centre VADER, Parc Valrose, 06000 Nice, France}
\address{RespInnovation SAS, 1300 route des crêtes, 06560 Sophia Antipolis, France}
\author{Barrett R. Mitchell}
\address{RespInnovation SAS, 1300 route des crêtes, 06560 Sophia Antipolis, France}
\author{Antonello Nicolini}
\address{Don Gnocchi Foundation, IRCCS, Via Alfonso Capecelatro, 66, 20148 Milano MI, Italia}
\author{Bernard Rousselet}
\address{Université Côte d'Azur, CNRS, LJAD, Centre VADER, Parc Valrose, 06000 Nice, France}
\author{Benjamin Mauroy$^{*,}$}
\email[Corresponding author: ]{benjamin.mauroy@univ-cotedazur.fr}
\address{Université Côte d'Azur, CNRS, LJAD, Centre VADER, Parc Valrose, 06000 Nice, France}

\begin{abstract}
Chest physiotherapy is a set of techniques used to help the draining of the mucus from the lung in pathological situations.
The choice of the techniques, and their adjustment to the patients or to the pathologies, remains as of today largely empirical.
High Frequency Chest Wall Oscillation (HFCWO) is one of these techniques, performed with a device that applies oscillating pressures on the chest.
However, there is no clear understanding of how HFCWO devices interact with the lung biomechanics.
Hence, we study idealised HFCWO manipulations applied to a mathematical and numerical model of the biomechanics of the lung.
The lung is represented by a fluid--structure interaction model based on an airway tree that is coupled to an homogeneous elastic medium. 
We show that our model is driven by two dimensionless numbers that drive the effect of the idealised HFCWO manipulation on the model of the lung.
Our model allow to analyze the stress applied to an idealised mucus by the air--mucus interaction and by the airway walls deformation.
This stress behaves as a buffer and has the effect of reducing the stress needed to overcome the idealised mucus yield stress.
Moreover, our model predicts the existence of an optimal range of the working frequencies of HFCWO.
This range is in agreement with the frequencies actually used by practitioners during HFCWO maneuvers. 
Finally, our model suggests that analyzing the mouth airflow during HFCWO maneuvers could allow to estimate the compliance and the hydrodynamic resistance of the lung of a patient.
\end{abstract}


\date{\today}

\maketitle


\twocolumngrid

\section{Introduction}

The human bronchial tree is a tree structure formed of about $200 \ 000$ bifurcating airways, whose sizes are decreasing at each bifurcation, resulting in a tree that is space-filling~\cite{weibel_morphometry_1963, weibel_pathway_1984, mauroy_optimal_2004}. 
The largest bronchus is the trachea that opens to the oesopharyngeal region and the smallest bronchi in the bronchial tree are the terminal bronchioles that open to the acini, where the exchange surface between alveolar air and blood is located.
Since the lung is connected to the ambiant air, it is susceptible to be in contact with external particles, potentially toxic or infectious.
Hence, the wall of almost all the airways of the bronchial tree~\cite{junqueira_junqueiras_2013} is covered with secretions --the pulmonary mucus-- that protect the lung.
The mucus captures the particles and is incessantly moved toward the oesopharyngeal region by the mucociliary clearance, a mechanism that moves the mucus thanks to cilia located on the bronchi walls.
 
 \begin{figure*}[t!]
\centering
\includegraphics[width=14cm]{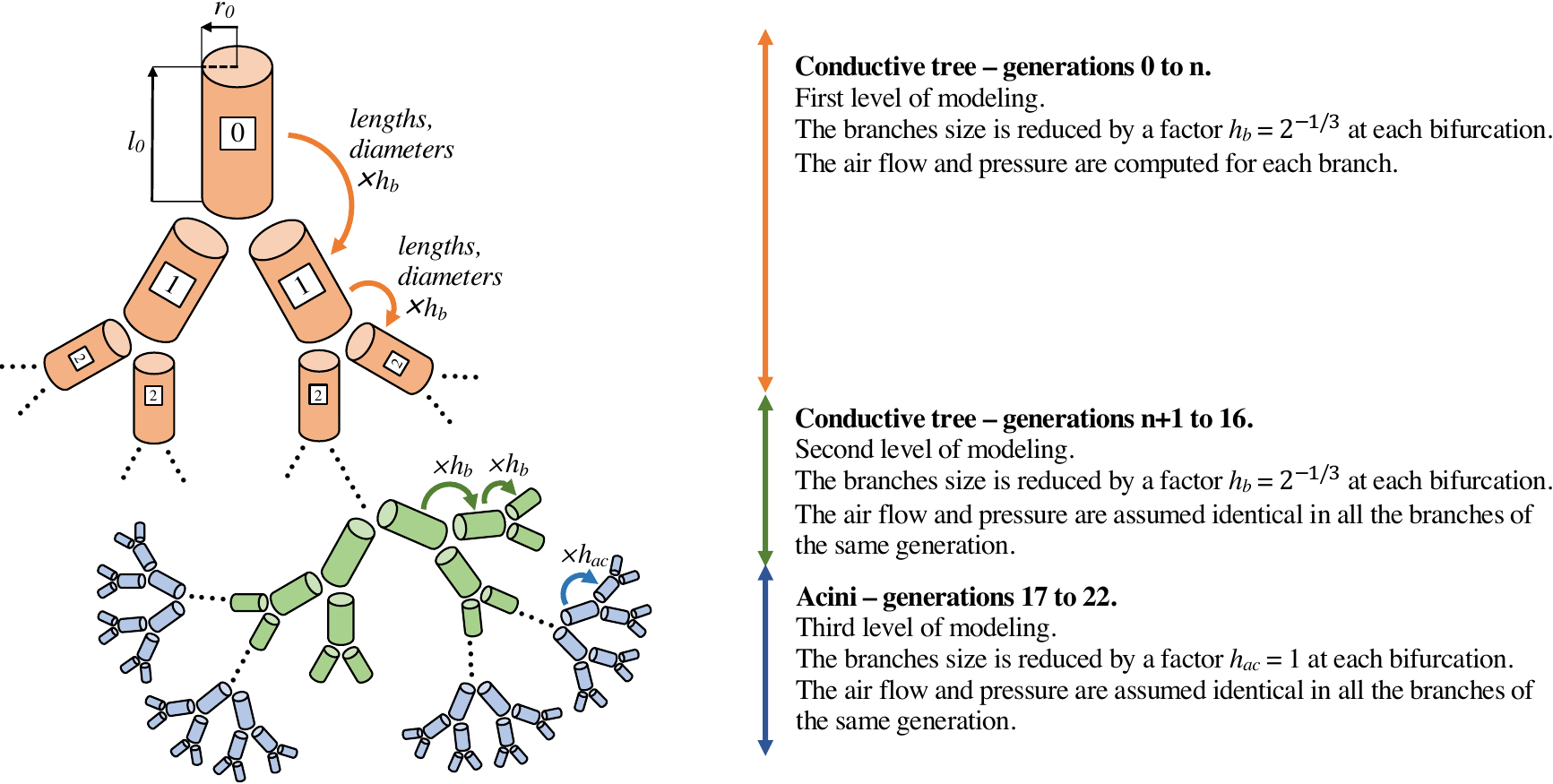}
\caption{The airway tree is modelled as a cascade of bifurcating cylinders representing the bronchi and the alveolar ducts. 
At each bifurcation, the size of the branches is decreasing by an homothetic factor, fixed to $h_b = \left( \frac12 \right)^{\frac13} \simeq 0.79$ in the conductive tree (17 first generations) \cite{weibel_morphometry_1963, mauroy_optimal_2004} and to $h_{ac} = 1$ in the acini (6 last generations) \cite{weibel_pathway_1984, tawhai_ct-based_2004}.
The number in the cylinders corresponds to the branches generation index, i.e. the number of bifurcations on the path between the root of the tree and the branch studied.
The first generation corresponds the root of the tree that mimics the trachea, its index is $0$.
The airway tree model decomposes into three levels: the first level corresponds to the first $n+1$ generations where the air flows and pressures are determined in each airway; the second level corresponds to the next $17-(n+1)$ generations where the air flows and pressures are assumed identical in all the airways belonging to a same generation; the third level corresponds to the acini (last six generations) where the air flows and pressures are also assumed identical in all the airways belonging to the same generation.}
\label{figTree}
\end{figure*}
 
Some pathologies are disrupting the mucociliary clearance and/or the cough. 
The mucociliary clearance can be altered by changes in the physical properties of the mucus~\cite{lai_micro-_2009, mauroy_toward_2011, mauroy_toward_2015} (viscosity, yield stress), by changes in the mucus production, as in cystic fibrosis~\cite{rubin_mucus_2007}, or by
perturbations of the cilia or of the cilia movement, as in primary ciliary dyskinesia or during bronchial inflammation in chronic obstructive pulmonary disease (COPD) or asthma~\cite{munkholm_mucociliary_2014}. 
Those diseases induce a stagnation of the mucus in the airways, increasing the risk of infections, and reduce the bronchi lumen area, hence altering the circulation of the air inside the bronchial tree. 
In such pathologies, therapeutical techniques are needed to help the patients to eliminate the excess of mucus and to recover, at least partially, their breathing capacity.

Chest physiotherapy is a common therapy used to compensate a defective mucociliary clearance or cough. 
It is based on mechanical forces applied on the thorax, aiming at changing the volume of the lung. 
This change of volume produces airflows that can potentially set the mucus in movement~\cite{mauroy_toward_2011, mauroy_toward_2015, stephano_modeling_2019, stephano_wall_2021}.
Chest physiotherapy can be performed manually by a practitioner or by the patient herself/himself --autogenic draining~\cite{agostini_autogenic_2007}.
The therapy can also be automated using specific mechanical devices. 
Many of these devices apply pressures in or on the lung to help the draining of the mucus, such as the Positive Expiratory Pressure technique (PEP), the Intrapulmonary Percussive Ventilation (IPV), the high frequency chest compression (HFCC) or the High Frequency Chest Wall Oscillations (HFCWO)~\cite{kluft_comparison_1996, dosman_high-frequency_2005, hristara-papadopoulou_current_2008, fornasa_characterization_2013, nicolini_effectiveness_2020}.
One of the challenge is to use the device that is the best adapted to the pathology or to the patient, and to determine its optimal functioning parameters in a framework where the knowledge of the therapeutic effects is mainly empirical and, hence, potentially controversial~\cite{mcilwaine_long-term_2013, mitchell_hfcwo_2013, fagevik_olsen_positive_2015, kuyrukluyildiz_what_2016, nicolini_safety_2018, reychler_intrapulmonary_2018}.

In this study, we will more particularly focus on HFCWO, which seems to be both efficient and well accepted by patients with specific pathologies~\cite{nicolini_effectiveness_2020}.
We define here HFCWO as the techniques that apply on the thorax~\cite{kluft_comparison_1996} small mechanical oscillations at relative high frequencies, i.e. a few Hertz to about twenty Hertz. 
Actually, HFCC is sometimes considered as part of HFCWO techniques although it applies an offset of "high" positive pressure to the small pressure oscillations~\cite{mcilwaine_physiotherapy_2006}.
In this work, we do not consider an offset of positive pressure, hence not we do not consider HFCC.

This work aims at characterizing, in an idealised framework, the biomechanics of the lung during HFCWO maneuvers.
We develop a mathematical and numerical model of the core biomechanical phenomena of the lung adapted to HFCWO and inspired from~\cite{berger_poroelastic_2016, pozin_tree-parenchyma_2017}.
{\cred The model is decomposed into a model of the air fluid dynamics in the bronchial tree and a model of the mechanics of the lung parenchyma.
The model of air flows the bronchial tree is the assembly of a cascade of three models of airway trees: one for the upper conductive tree, one for the lower conductive tree and one for the acini.
This decomposition allows to account, depending on the scale, for the different geometries of the airways and for the different regimes of air circulation.
Also, adjusting the size of the three levels allows to tune the bronchial tree model complexity \cite{clark_capturing_2017}.
The model of the mechanics of the lung parenchyma is based on linear elasticy.}
Then we apply to that model of the lung idealised HFCWO maneuvers.


	\section{Model of the lung}
	\label{S:2}
		
We assume that the lung at functional residual capacity fills a domain $\Omega$ of the 3D space. 
We consider the lung as two regions with different physics that are interacting together~\cite{berger_poroelastic_2016, pozin_tree-parenchyma_2017,donovan_systems-level_2016, birzle_viscoelastic_2019}. 
The first region, called the tree region, corresponds to the airways and alveolar ducts.
The second region, called the tissue region, corresponds to the lung's parenchyma.\\
		
\noindent{\bf The tree region.} 
Different frameworks have been used in the literature to model the bronchial tree, from the most complex, based on 3D geometries that are reconstructed from CT-scans of the lung~\cite{tawhai_ct-based_2004, rosell_three-stage_2013, kim_effect_2018, longest_use_2019}, to idealised tree geometries.
Idealised tree geometries allow to develop more tractable models.
They are either generated by algorithms that mimic the statistics of the airways~\cite{kitaoka_three-dimensional_1999, tawhai_generation_2000, choi_1D_2019} or by using data-based models, with different levels of complexity, going from fractal-like models (one or two parameters)~\cite{weibel_morphometry_1963, mauroy_optimal_2004, mauroy_influence_2010, guha_finding_2016, tsega_computational_2018} to more complex geometries where each level of bronchi is described independently~\cite{lambert_computational_1982, mauroy_optimal_2008, mauroy_toward_2015,  stephano_wall_2021}.

Here, the airway tree is represented by three different modelling levels, see Figure \ref{figTree}.

The upper conductive airways are modelled by rigid cylinders assembled into a bifurcating tree that mimics the structure of the bronchial airways. 
The size of the cylinders is decreasing at each bifurcation with a constant ratio $h_b=\left(\frac12\right)^{\frac13} \simeq 0.79$ \cite{weibel_morphometry_1963, mauroy_optimal_2004}.
The generation index of a cylinder in the tree corresponds to the number of bifurcations between the root of the tree and that cylinder.
The root of the tree mimics the trachea and corresponds to the first generation with index $0$. 
In this model, all the branches in the same generation have the same geometrical properties, but their inner air fluid dynamics can be different.
The first level of the tree corresponds to $n+1$ successive generations.  
The total number of terminal branches is $N=2^{n}$.

The number of generations for the first level is $n+1$ and it can be lower than the approximate average of $17$ generations of the conductive airways \cite{weibel_pathway_1984}. 
Hence, the second modelling level mimics the $17-(n+1)$ generations of conductive airways.
This level corresponds to a set of subtrees, connected by set of two at each terminal branches of the tree of the first modelling level.
The subtrees geometry is similar to the geometry of the tree of the first level.
However, within one of these subtrees, we assume that the air physics is identical in all the airways with the same generation index.
The total number of terminal branches of the second modelling level is $2^{16}$.

Finally, the third modelling level mimics the acini.
An acinus can be viewed as a $6$ generations dichotomous subtree with rigid cylindrical branches.
In the acinus, we can consider that the size of the branches remain the same at each bifurcation, i.e. the size reduction ratio between two successive generations is $h_{ac} = 1$ \cite{haefeli-bleuer_morphometry_1988}.
The third modelling level corresponds to two acinus models connected to each terminal branches of the second modelling level.\\

\begin{figure*}[t!]
\center
\includegraphics[height=5cm]{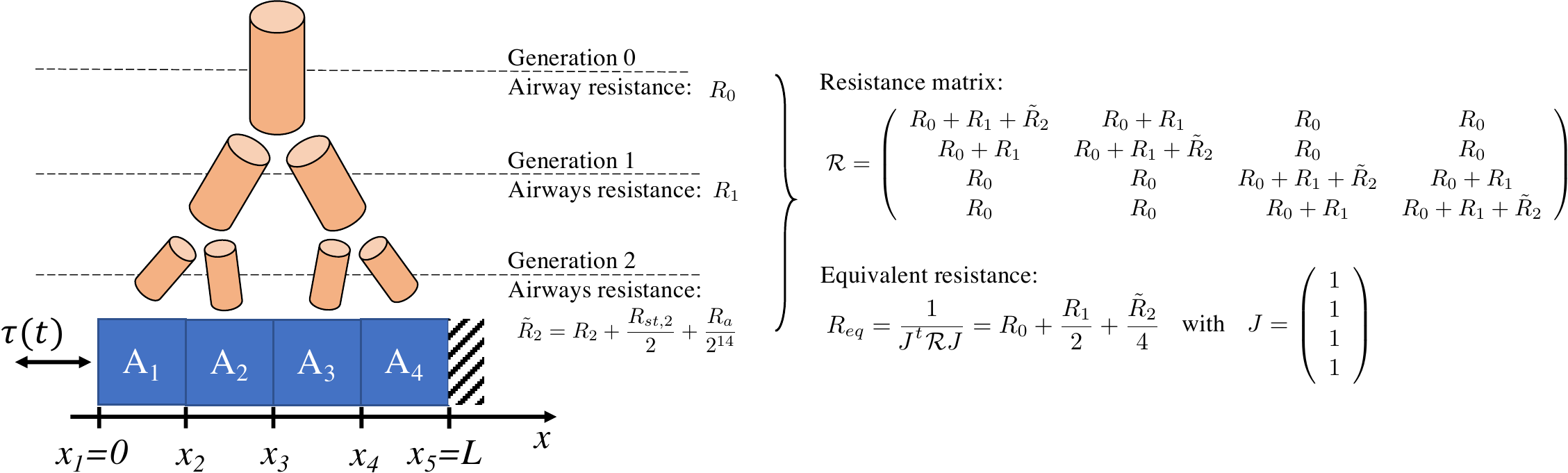}
\caption{Example of the model for $n=2$.
In that case, the first modelling level is made of three generations and four terminal branches.
{\cred All the tree branches of a same generation $i$ are identical and have the same hydrodynamic resistance $R_i$.}
Each terminal branch is coupled to one of the four subregions $(A_i)_{i=1,2,3,4}$ of the 1D tissue region $\Omega = [0,L] = \cup_{i=1}^N A_i$.
{\cred The hydrodynamic resistance $R_2$ of each terminal branch is replaced with a modified resistance $\tilde{R}_2$, which accounts for the hydrodynamic resistances of the subtrees connected to that terminal branch.
The resistance $R_{st,2}$ corresponds to the resistance of one of the two subtrees that are connected to each terminal branch.
Each of these subtrees is part of the generations $3$ to $17$.
The resistance $R_a$ corresponds to the resistance of one of the $2^{14}$ acini connected to each terminal branch of generation $2$.
The acini forms the last sixth generations of the tree.}
An oscillating pressure $\tau(t)$ is applied at $x=0$.
The material is fixed at $x=L$.
The rate of volume change of a subdomain $A_i$ corresponds to the air flow going through the corresponding terminal branch.
The resistance matrix $\mR$ and the associated equivalent hydrodynamic resistance $R_{eq}$ corresponding to the pictured case are written on the right of the figure.
}
\label{F:2}
\end{figure*}

\noindent{\bf Pressure--flow relationship.} 
The air in the branches is considered as an incompressible Newtonian fluid with viscosity $\eta$.
We neglect the influence of the bifurcations on the air flow and the pressure drop in the nasopharyngeal pathway \cite{maury_respiratory_2013}. 
The reference pressure is the atmospheric pressure. 

{\cred The steady-state Poiseuille's regime corresponds to a low air flow, fully developed and axisymmetric, where the acceleration of the fluid is neglected. 
In this regime, the air flow $f$ in a cylindrical airway is related to the inlet pressure $p_{in}$ and the outlet pressure $p_{out}$ by
$$
p_{in} - p_{out} = R \, f
$$ 
where $R=\frac{8 \eta l}{\pi r^4}$ is the hydrodynamic resistance of the cylinder, with $r$ and $l$ the respective radius and diameter of the cylinder. 
The radii and lengths of the branches in the generation $i$ follow a scaling laws $r_i=h_b^i r_0$ and $l_i=h_b^i l_0$ with $r_0$ and $l_0$ the radius and length of generation $0$. 
Consequently, the Poiseuille hydrodynamic resistance also follows a scaling law $R_i = R_{i-1}/h_b^3$ ($i>1$), with $R_0$ the resistance of the root of the tree. 
As $h_b = \left(\frac12\right)^{\frac13}$, $R_i = 2^i R_0$ ($i=0,\dots,n$). 
However, the Poiseuille regime does not account for the complex flow patterns occurring in the large airways that affects the airflows and pressures distribution in the tree \cite{faizal_computational_2020}.
Hence, we use here a "corrected" Poiseuille regime by multiplying the hydrodynamic resistances of the airways in the first levels of the tree (generations $0$ to $n$) by an ad-hoc factor $c$.
The factor $c$ will be determined using a calibration process detailed in Appendix \ref{restApp}.
}

From the pressure--flow relationships in each branch, we can derive a global linear relationship for the whole tree. 
We define the flows vector $F = (f_j)_{j=1,\dots,N}$, with $f_j$ the air flow at the $j$-th terminal branch and the pressures vector $P = (p_j)_{j=1,\dots,N}$, with $p_j$ the air pressure at the $j$-th terminal branch. 
The linear relationship between the pressures and flows vectors is based on the resistance matrix $\Rm = (\Rm_{ij})_{ij=1,\dots,N}$ of the airway tree \cite{grandmont_viscoelastic_2006, dubois_de_la_sabloniere_shape_2011},
\begin{equation}
\label{eq1}
P = \Rm F    
\end{equation}
The coefficients of the resistance matrix $\Rm$ are sums of the hydrodynamic resistances of the cylinders in the paths and the subpaths linking the root of the tree and the terminal branches of the tree.
Moreover, the equivalent resistance $R_{eq}$ of the tree relates an identical pressure $p$ applied at each terminal branch with the total amount of airflow in the tree $F_T$ (i.e. the air flow in the first generation), $p = R_{eq} F_T$.
The equivalent resistance can be computed from the resistance matrix \cite{mauroy_optimal_2008} $\Rm$ by $R_{eq} = (J^t \Rm^{-1} J)^{-1}$ with $J = (1,\dots,1)^t \in \mathbb{R}^N$.
An example with $n=2$ is given in Figure \ref{F:2}.

In order to account for the influence of the subtrees of the second and third modelling levels, the resistances of the terminal branches of the tree of the first modelling levels are modified.
Since the physics of air in the second and third modelling levels are assumed identical per generation of subtrees, the pressures at the terminal branches of a single subtree are all the same.
Hence, each subtree hydrodynamic response is determined based on its equivalent hydrodynamic resistance only. 
The hydrodynamic resistance of one subtree of the second modelling level is $R_{st,n} = \frac{R_{n}}{h_b^3} \sum_{i=0}^{17-n-2} \left(\frac{1}{2 h_b^3}\right)^i = (17 - (n + 1)) \frac{R_{n}}{h_b^3}$ and for the third modelling level, it is $R_{a} = R_{16} \sum_{i=0}^{5} \left(\frac{1}{2 h_{ac}^3}\right)^i$.
To each terminal branch of the first modelling level of the tree are connected two subtrees of the second modelling level and $2^{17-(n+1)}$ subtrees of the third modelling level.
Finally, the resistance $R_{n+1}$ of the terminal branches of the tree of the first modelling level is replaced by the resistance $\tilde{R}_{n+1}$ that accounts for the subtrees, 
\begin{equation}
\tilde{R}_{n} = R_{n} +  \frac{R_{st,n}}{2} + \frac{R_{a}}{2^{17-(n+1)}}
\end{equation}\\

\begin{table*}[t!]
\centering
\begin{tabular}{lcl}
\hline
\multicolumn{3}{c}{\bf Model input parameters}\\
\hline
\multicolumn{3}{c}{}\\
\bf Physical quantity & \bf Parameter name & \bf Value\\ 
\multicolumn{3}{c}{}\\
Tree root radius (trachea radius) \vspace{0.15cm} & $r_0$ & $1$ cm~\cite{weibel_pathway_1984}\\
Tree root length (reduced trachea length) \vspace{0.15cm} & $l_0$ & $6$ cm~\cite{weibel_pathway_1984} and Appendix \ref{S:3:2:1}\\
Lung characteristic size (human, adult) \vspace{0.15cm} & $L$ & $20$ cm~\cite{weibel_pathway_1984}\\
Lamé parameters of the tissue region (1D) \vspace{0.15cm} & $\lambda + 2 \mu$ & $2700$ Pa~\cite{pozin_tree-parenchyma_2017},  Appendix \ref{restApp}\\
Lung density \vspace{0.15cm} & $\rho$ & $100$ kg.m$^{-3}$~\cite{pozin_tree-parenchyma_2017}\\
Resistance matrix of the airway tree \vspace{0.15cm} & $\mR$ & cmH$_2$O.L$^{-1}$.s, see \cite{mauroy_optimal_2008, grandmont_viscoelastic_2006, dubois_de_la_sabloniere_shape_2011}\\
Hydrodynamic resistance of the airway tree \vspace{0.15cm} & $R_{eq}$ & $1.0$ cmH$_2$O.L$^{-1}$.s\cite{maury_respiratory_2013}, Appendix \ref{restApp} \\
Bronchi walls Young's modulus \vspace{0.15cm} & $E_b$ & $6250$ Pa \cite{mauroy_optimal_2008}, Appendix \ref{airwayRadius}\\
Mucus Young's modulus \vspace{0.15cm} & $E_m$ & $1.0$ Pa \cite{lai_micro-_2009}, Appendix \ref{stressMucus}\\
Idealized HFCWO frequency \vspace{0.15cm} & $f = 1/T$ & range $1$ -- $18$ Hz~\cite{nicolini_effectiveness_2020}\\
Idealized HFCWO applied pressure \vspace{0.15cm} & $A$ & $200$ Pa (computed)\\
\multicolumn{3}{c}{}\\
\hline
\multicolumn{3}{c}{\bf Characteristic quantities}\\
\hline
\multicolumn{3}{c}{}\\
\bf Physical quantity & \bf Variable name & \bf Expression\\ 
\multicolumn{3}{c}{}\\
Time \vspace{0.15cm} & $T$ & $\frac1f$\\
Velocity \vspace{0.15cm} & $v$ & $\frac{L}{T}$\\
Wave velocity \vspace{0.15cm} & $c$ & $\sqrt{\frac{\lm}{\rho}}$\\
Displacement \vspace{0.15cm} & $\Upsilon$ & $\frac{A L}{\lm}$\\
Equivalent resistance of the tree \vspace{0.15cm} & $R_{eq}$ & $\frac1{J^t \mR J}$\\
Air pressure in the tree \vspace{0.15cm} & $p_L$ & $R_{eq} S_L v$\\
Effective air pressure in the tree \vspace{0.15cm} & $\mP$ & $\frac{A}{\lm} p_L$\\
Tissue inertia & -- & $\rho v^2$\\
\multicolumn{3}{c}{}\\
\hline
\multicolumn{3}{c}{\bf Dimensionless numbers}\\
\hline
\multicolumn{3}{c}{}\\
\bf Name & \bf Variable name & \bf Expression\\ 
\multicolumn{3}{c}{}\\
Euler number  \vspace{0.1cm} & $\mE$ & $\frac{p_L}{\rho v^2}$\\
Inverse of Cauchy number \vspace{0.2cm} & $\mB = \frac{1}{\mC}$ & $\left( \frac{c}{v} \right)^2$\\
Lung Mechanics number \vspace{0.1cm} & $\mL_M = \frac{\mathcal{E}}{\mathcal{B}}$ & $\frac{p_L}{\lm}$\\
\multicolumn{3}{c}{}\\
\hline
\end{tabular}
\caption{Input parameters, characteristic quantities and dimensionless numbers used in this work. 
}
\label{params}
\label{T:1}
\end{table*}

\noindent{\bf Influence of the air on the tissue region.}
The domain $\Omega$ reflects the lung's spatial occupation and is decomposed into $N = 2^{n}$ regions $(A_i)_{i=1..N}$.
Each $A_i$ is fed by a single terminal branch of the tree, as schematized in Figure \ref{F:2}.
We neglect the volumetric influence of the bronchial tree in the $A_i$'s, since it represents a small fraction of the volume of the lung, about $10\%$~\cite{weibel_pathway_1984}. 
We assume that the lung tissue behaves as an homogeneous elastic material~\cite{werner_patient-specific_2009, berger_poroelastic_2016, pozin_tree-parenchyma_2017} and we assume small strains theory. 
This choice is well adapted to HFCWO, since this technique applies small oscillating pressures only. 

Lung's tissue displacements at location $x \in \Omega$ and at time $t \in  \mathbb{R}_{+}$ are represented by the variable $u(x,t) \in \mathbb{R}^m$. 
The general displacements equations are
\begin{equation}
\label{eq2}
\rho \frac{\partial^2 u}{\partial t^2} - \rm{div}(\sigma(u)) = 0
\end{equation}
where $\rho$ is the volumetric mass density of the material and $\sigma(u)$ is the stress tensor.
The boundary $\partial \Omega$ of $\Omega$ is decomposed into two regions: $\Gamma_1$ represents the region where the stress is applied and $\Gamma_2$ represents the region where there is no displacement. 
Hence, the boundary conditions on $\partial \Omega$ and the initial conditions in $\Omega$ are
\begin{equation}
\label{eq3}
\left\{
\begin{array}{ll}
\sigma(u).n = \tau(x,t) & \text{ $x \in \Gamma_1$}\\
u(x,t) = 0 & \text{ $x \in \Gamma_2$} \\
u(x,0) = u_0(x) & \text{ for $x \in \Omega$}
\end{array}
\right.
\end{equation}
The quantity $\tau$ is the pressure applied on the boundary and is the source of the system dynamics.
We assume the material to be isotropic and to be linear elastic. 
The elastic stress $\sigma_{\rm{e}}(u)$ relates to the displacement $u$ as
$
\sigma_{\rm{e}}(u) = \lambda \; \rm{tr}\left(\epsilon(u)\right)I+2\mu\; \epsilon(u)
$
, with $I$ the identity matrix, $\epsilon(u) = \frac12 \left( \nabla u + \nabla u^t \right)$ and $\lambda$ and $\mu$ the Lamé parameters.

The air flowing out of the exchange surface goes through the bronchial tree.
Any change of the volume of the material is counteracted by the resistance to the air flow induced by the tree structure. 
This is reflected in the material stress--strain relationship by a supplementary local stress, actually a pressure, that depends on how the air is conveyed in the tree. 
Each terminal branch $i$ induces an homogeneous pressure $p_i$ in its corresponding region $A_i$. 
The pressures are determined by the rate of volume change of the $A_i$ along time. 
In the case of small deformations, this rate, which corresponds to the air flow, can be approximated with \cite{pozin_tree-parenchyma_2017}
\begin{equation}
\mathcal{F}_i[\dot{u}]=\int_{A_i} -{\rm{div}} (\dot{u}) dx
\label{eq3b}
\end{equation}
where $\dot{u} = \frac{\partial u}{\partial t}$.
We denote $\mathcal{F}[\du] = \left(\mathcal{F}_i[\du]\right)_{i = 1,\dots,N}$ the vector of air flows at the terminal branches. 
The pressure $p_i$ in one $A_i$ depends on the air flows $\mathcal{F}[\du]$ in all the terminal branches, see equation (\ref{eq1}). 
Hence,  
the pressure $p_{\rm{tree}}(\mathcal{F}[\du])$ induced by the air in the material is a piecewise function,
$p_{\rm{tree}}(\mathcal{F}[\du])(x) = p_i(\mathcal{F}[\dot{u}]) = \left(\Rm \mathcal{F}[\du]\right)_i$ for $x \in A_i$.
Hence, the inner stress tensor induced by the tree is $\sigma_{\rm tree}(\du) = - p_{\rm{tree}}(\mathcal{F}[\du])I$.
This stress is not continuous at the boundaries of the $A_i$.

Finally, the stress--strain relationship for the model of the tissue region is 
\begin{equation}
\label{eq5}
    \mathbf{\sigma}(u,\mathcal{F}[\dot{u}])=\underbrace{\lambda { \rm Tr}(\epsilon(u))I+2\mu\epsilon(u)}_{\sigma_{\rm{e}}(u)}  \underbrace{- \ p_{\rm{tree}}(\mathcal{F}[\dot{u}])I}_{\sigma_{\rm tree}(\dot{u})}
\end{equation}
The resulting stress--strain relationships in equation (\ref{eq5}) is that of a viscoelastic material, with a non-local viscous behavior.
Due to the discontinuity of $\sigma_{\rm tree}$, the correct mathematical way to express the system equations is the weak form, see details in Appendix \ref{weakApp}.

We will consider the material to be able to deform only in the direction $x$ and to be rigid in the two other directions, with a constant cross-section $S_L = L^2$.
Under these conditions, the equations become unidimensional in space on the domain $\Omega=[0,L]$.
We assume that the pressure $\tau$ is applied at $x=0$ and that the material is fixed at $x=L$. 
From now on, the applied pressure $\tau$ is assumed sinusoidal in time, i.e. $\tau(t) = A \sin(\frac{2 \pi}{T} t)$ with $A$ the amplitude of the applied pressure and $T$ its period.

\section{Results}
\label{S:3}

\noindent{\bf Physical analysis.} To reach a better understanding of the equations and to determine the intrinsic parameters of the problem, the equations are rewritten using a dimensionless formulation. 
The space, the time and the amplitude of the solution are adimensionalized with $y = x/L$, $s = t/T$, $u(x,t) = \Upsilon\tilde{u}(y,s) = \Upsilon\tilde{u}(\frac{x}{L},\frac{t}{T})$, $p_i(\mathcal{F}[\du])=\mathcal{P} \tp_i(\tilde{\mathcal{F}}[\frac{\partial \tilde{u}}{\partial s}])$, $\ttau(s) = \tau(t)/A$.
The quantities $L$ represents the characteristic size of $\Omega = [0,L]$ and the space domain becomes $\tilde{\Omega}=[0,1]$.
The quantity $T$ is the system characteristic time, given by the period of the applied pressure $\tau(t)$.
The dimensionless formulation is derived in Appendix \ref{dimApp}. 
The quantity $\Upsilon = \frac{A L}{\lm}$ represents the characteristic displacement of the structure and $\mathcal{P} = R_{eq} \frac{S_L \Upsilon}{T}$ the characteristic pressure. 
$\tilde{\Omega}$ is decomposed into $N$ subsets $\tilde{A_i}$, which are the transformations by the adimensionalization of the corresponding $A_i$ in the original space. 
We define the characteristic velocity $v$ to cross the whole system in a time $T$ as $v = L/T$.

With these new variables, the dimensionless energetic balance, computed in Appendix \ref{S:3:2:2}, is
\begin{equation}
\begin{aligned}
\frac{d}{ds} & \overbrace{\left(\frac12 \int_0^1 \left(\frac{\partial \tilde{u}}{\partial s}(y,s)\right)^2 dy\right.}^{\text{kinetic energy}}  
+ \overbrace{\left.\mathcal{B} \frac12 \int_0^1 \left(\frac{\partial  \tilde{u}}{\partial y}(y,s)\right)^2 dy\right)}^{\text{elastic energy}}\\
&= \underbrace{\mB \ttau(s) J^t \tilde{\mathcal{F}}[\frac{\partial  \tilde{u}}{\partial s}]}_{\text{input power}}
+ \underbrace{\mathcal{E} \ \sum_{i=1}^N \tp_i(\tilde{\mathcal{F}}[\frac{\partial  \tilde{u}}{\partial s}]) \tilde{\mathcal{F}}_i[\frac{\partial \tilde{u}}{\partial s}]}_{\text{dissipated viscous power}}
\end{aligned}
\label{eq9}
\end{equation}
The number $\mathcal{B}= \lm / \rho v^2$ is the inverse of the system Cauchy number.
It compares the elastic forces in the material with the inertial forces. 
The number $\mathcal{E} = \frac{p_L}{\rho v^2}$ is the Euler number of the system.
The pressure $p_L = R_{eq} S_L v$ represents the non-coupled characteristic pressure in the terminal branches of the tree, i.e. in the absence of the coupling with the respiratory zone. 
In comparison, the pressure $\mP = \frac{A}{\lm} p_L$ represents the efficace characteristic pressure resulting from the coupling. 
The Euler number of the system compares the pressure forces induced by the viscous dissipation of the air flow in the bronchial tree with the inertial forces in the material.

Finally, the system is characterized by the two dimensionless numbers $\mathcal{B}$ and $\mathcal{E}$. 
Their ratio $\mL_M = \mE / \mB$ is called the Lung Mechanics number, it compares the elastic energy to the dissipation.
When $\mL_M << 1$, the system behaves as a wave equation and the evolution of the system total energy depends on the boundary condition in $y=0$, and more precisely on $\ttau(s)$.
Additionally, if $\mathcal{B}>>1$, then the energy is mainly stored as elastic energy and the wave propagates rapidly. 
The material displacement is close to a Laplacian (or diffusive) profile, i.e. linear in 1D.
If $\mathcal{B}<<1$, then the energy is mainly stored as kinetic energy and the wave propagates slowly.

On the contrary, when $\mL_M >> 1$, then the system is quickly damped.
Hence, the kinetic energy, the elastic energy and the airflows quickly drop to zero.
Details about the influence of the tree structure on the tissue region dynamics is given in Appendix \ref{S:3:2}.\\

 \begin{figure*}[p!]
		\centering		
		{\includegraphics[width = 12cm]{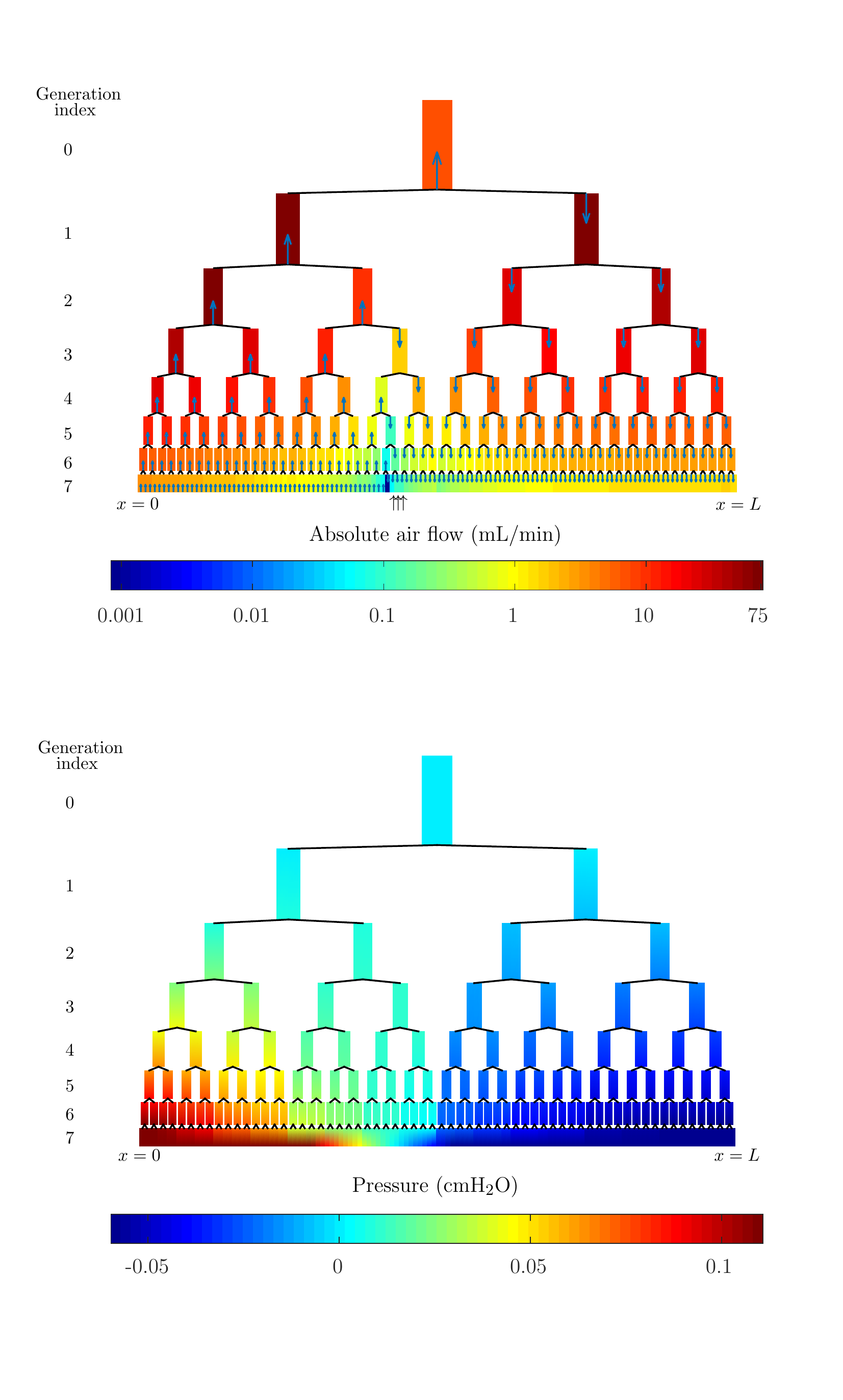}}
		\caption{\cred Air properties in the first level model of the bronchial tree with $8$ generations ($n=7$): mean airflows (up) and mean absolute pressures (down).
		The data are plotted for an idealized HFCWO frequency of $20$ Hz at the beginning of a HFCWO cycle. 
		This time corresponds to a maximal time derivative of the sinusoidal pressure applied at the position $x=0$ of the tissue (on the left).
		The rectangles represent the cylindrical airways to scale: their widths correspond to the airways diameters and their lengths to the airways lengths.
		The blue arrows on the upper plot represents the air flow orientation in the airways.
		The branches with generation index $7$ account for the airways of the deeper generations, as schematized in Figure \ref{figTree}.			
		The air circulates between different parts of the tree, getting away from the parts that are connected to the compressed regions of the tissue, where the air pressure is higher (left part of the tree in the example plotted).
		The air is either expelled through the root of the tree (trachea) or sent to the parts of the tree connected to the regions of the tissue with lower pressures (right part of the tree in the example plotted).
		The three black arrows in the upper plot indicate the terminal branches for which the pressure in the tissue is positive although the tissue is expanding.
		This phenomena occurs at the propagation front of the wave, where the material is in transition between expansion and compression.
		}
		\label{F:6Tree}
\end{figure*}

\noindent{\bf Application to HFCWO.}  Approximated solutions of the model equations are obtained using the finite elements method, implemented in the open source software {\it Octave} \cite{eaton_gnu_2019} and available in \cite{brunengo_code_2021}.
Our algorithm is validated by comparison with unidimensional analytical solutions, see Appendix \ref{numSim}.
The physiological data used for the input parameters of the model are given in Table \ref{T:1}.
The ventilation at rest in human is thoroughly studied in the literature, hence it is used to calibrate and validate our model, see Appendix \ref{S:4:1}.
{\cred The calibration consists in adjusting the hydrodynamic resistance correction factor $c$ of the airways in the tree first level of modeling.
As detailed previously, this factor compensates for geometrical and fluid dynamics features neglected in our model.
The value of $c$ is calibrated so that the equivalent hydrodynamic resistance of our model of the bronchial tree is compatible with the physiology.
As no data is available for HFCWO in the literature, the calibration and validation in the case of HFCWO is not possible. 
Hence, we use rest ventilation to calibrate and validate our model and assume that the configuration remain relevant for HFCWO. 
With a value $c=20$, the tree equivalent hydrodynamic resistance is $R_{eq} \simeq 1$ cmH$_2$O.L$^{-1}$.s, in accordance with the data for healthy adult lungs \cite{maury_respiratory_2013}.
The model is then validated by comparing its predictions at rest regime for tidal volume, mouth airflows and acinar air pressures with the physiological data available in the literature \cite{weibel_pathway_1984}.}
Once calibrated and validated, our model is used to mimic HFCWO manipulation.

The amplitude $A$ and the period $T$ of the boundary condition at $x=0$ are adjusted to mimic an idealised HFCWO maneuver.
We consider HFCWO to work as an applied sinusoidal pressure $\tau(t)=A \sin (2 \pi t / T )$, and we denote $f = 1/T$ the frequency. 
The typical frequencies used in HFCWO device are in the range $1$ Hz to $20$ Hz. 
To our knowledge, the amplitude of the force felt by the lung due to the pressure on the thorax is not documented yet.
Since our model is linear in $A$, we can easily determine the solution for any value of $A$ from a single computation once the other parameters, such as the frequency, have been fixed.
Our goal is to compare the efficiency of the different frequencies by observing the airflow induced by HFCWO.
We consider that a HFCWO device is more efficient if the airflow is larger.
Notice that due to the linearity of the equations relatively to the boundary condition at $x=0$, mixing the rest ventilation and HFCWO would bring an amount of airflow that would be the sum of the airflows induced by the ventilation and by the HFCWO computed separately.
Hence, in order to isolate the effects of HFCWO in our simulations, we do not account here for the lung's ventilation.\\

\cred\noindent{\bf Air--tissue interactions.}
The tissue and air mechanics are affecting each other.
When the tissue undergoes a compression, the air is going from the tissue into the airway tree; when the tissue undergoes an expansion, the air is going from the tree into the tissue. 
The flow of air through the tree is conserved and air is exchanged with ambient air through the first generation airway, which mimics the trachea, and with all the terminal branches connected to the tissue.
The flow of air distributes in the tree depending on the air pressures distribution, which results from the hydrodynamic resistances of the pathways between the terminal branches and between the terminal branches and the trachea.
The air pressures at the end of the terminal branches are felt by the tissue and, in turn, affect its propension to compress or expand.

This behavior leads to complex flow patterns in the tree, as shown in Figure \ref{F:6Tree}, where the air flows and pressures in the tree are represented at the beginning of a HFCWO cycle ($A=200$ Pa, $f=20$ Hz).
In this example, the deformation wave induces the compression of the left part of the tissue, where air is flowing from the tissue into the corresponding terminal branches.
Because the air is pushed into a resistive tree, its pressure in this part of the tissue tends to be positive and opposes to the compression of the tissue.
The air flow is then directed toward the trachea, where it reaches out to ambient air, and towards other terminal branches, which are connected to the tissue parts that are undergoing an expansion and have lower pressures.
If the low pressure in these regions is negative, it opposes to the expansion of the tissue.
Depending on the hydrodynamic resistance distribution and air flow distribution, it is possible to have compression with negative pressure or expansion with positive pressure.
In these cases, the air pressure does not oppose the tissue deformation but instead favors it. 
Expansion with positive pressure occurs at the propagation front of the wave, where the tissue is in transition between tissue expansion and compression.
Compression with negative pressure occurs at the propagation queue of the wave, where the tissue is in transition between compression and expansion. 
The patterns can be even more complex in trees with asymmetrical bifurcations, where the air--tissue interactions are affected by the specificity of the geometry of the tree.
Such an example, based on a tree with physiological branches size and asymmetric bifurcations, is presented in Appendix \ref{RaabeTree}.\\ \cb

\noindent{\bf An optimal range of frequencies.}
HFCWO is known to help move the mucus by affecting its rheology --out of the scope of this study-- and by applying stresses in the mucus, either by the air--mucus interactions~\cite{freitag_mobilization_1989, mauroy_toward_2011, mauroy_toward_2015, stephano_wall_2021} or by the mechanical deformations induced by the oscillations of the airways walls.
The air volume $V_p$ exchanged with the ambiant air and the airflows created by HFCWO are dependant on its working frequency, whose recommended values are based on empirical knowledge.
Hence, we study with our model the influence of the HFCWO frequencies on the inhaled air volume and on the tracheal airflow (flow in the first generation of the tree).
We assume that the amplitude of the applied pressure $A$ at the boundary $x=0$ is fixed to $A = 200$ Pa and make the frequency of HFCWO range between $1$ Hz and $18$ Hz.

 \begin{figure*}[t!]
		\centering		
		{\includegraphics[width = 14cm]{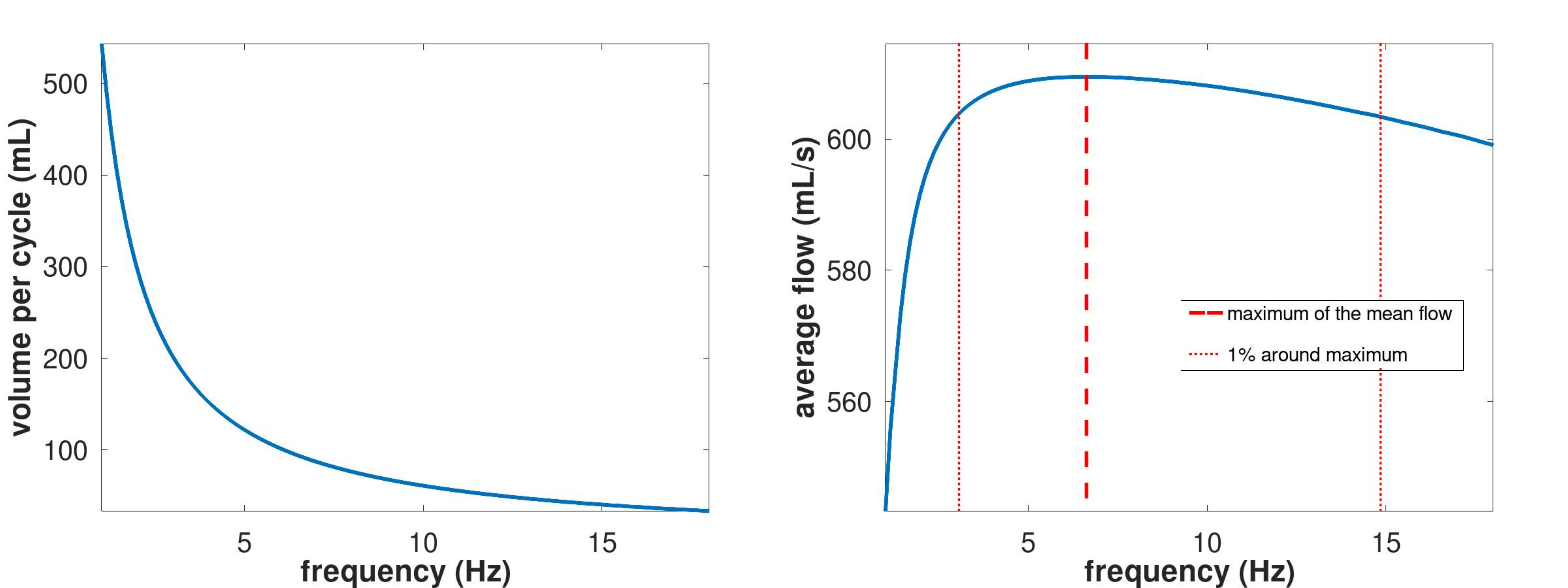}}
		\caption{{\bf Left:} Volume of air (mL) ventilated at each cycle of the applied constraint versus the frequency of the applied pressure. 
		{\bf Right:} Average total air flow in the tree (mouth airflow, mL/s) versus the frequency of the applied pressure. 
		The long red long-dashed line represents the maximum of the average total air flow, reached for a frequency of about $6.5$ Hz.
		The red short-dashed line represents the range of frequencies for which the average total air flow is within a range of 1\% of the maximum.
		}
		\label{F:5}
\end{figure*}

The model predicts that $V_p$ decreases as the frequency increases, with a decreasing slope, as shown in Figure \ref{F:5} (left). 
Also, the average airflow increases for frequencies lower than $6.5$ Hz and then decreases. 
Hence, the maximal airflow is reached at an optimal frequency $f_o = 6.5$ Hz, see Figure \ref{F:5} (right). 
As our model is linear in the amplitude of the applied pressure $A$, $A$ affects only the amplitude of the volumes and of the airflows, but not the location of the maximum.
At the optimal frequency, the two dimensionless numbers are of the same order of magnitude with $\mathcal{B} = 16.56$ and $\mathcal{E} = 32.74$.
The acceleration (1) has a low influence on the system relatively to the elasticity ($\mB$) and to the dissipation ($\mE$). 
The value of the Lung Mechanics number $\mL_M = \mE / \mB$ is $1.97$ and the dissipation affects slightly more the system than the elasticity.
In comparison, during rest ventilation $\mathcal{B} \simeq 17\,000$ and $\mathcal{E} \simeq 1\,000$, see Appendix \ref{restApp}, and dissipation plays a smaller role as $\mL_M = 0.062$.
Hence at rest, a significant portion of the elastic power developed during inspiration is stored and can be recovered during expiration \cite{weibel_pathway_1984}.
In comparison, during HFCWO a larger fraction of the power put in the system is lost to dissipation.

Near the optimal frequency $f_o$, the airflow is actually on a plateau. 
In the range of frequencies from $3$ Hz to $15$ Hz, the amount of airflow remains within $1\%$ of the maximum.
By maximizing the airflow in the tree, this range of frequencies maximizes the global displacements of the material and the air--mucus interactions.

The optimal frequency $f_o = 6.5$ Hz, which maximizes the average air flow, corresponds to the fundamental frequency of the system without the damping influence of the tree, i.e. $f_1 = \sqrt{\frac{\lm}{\rho}}\frac{1}{4 L} = 6.5$ Hz \cite{rao_vibration_2007}.
\cred
This result can be highlighted using a parallel with damped oscillators with a single degree of freedom.
The displacement $z(t)$ of such oscillators follows an equation of the form~\cite{harris_harris_2002} $m \ddot{z} - \zeta \dot{z} - k z(t) = a \sin(\omega t)$, with $m$ the mass of the oscillator, $\zeta$ the damping coefficient, $k$ the spring coefficient and $a \sin(\omega t)$ an external oscillating force of amplitude $a$ and frequency $\omega$ applied to the oscillator.
In this case, the amplitude of the velocity $v = \dot{z}$ of the oscillator is maximal when the frequency of the applied force equals the velocity resonance frequency $\omega_v$.
The velocity resonance frequency is independent of the damping and equal to the fundamental frequency of the oscillator~\cite{harris_harris_2002}, $\omega_v = \sqrt{k/m}$.
In our model, the airflow is directly related to the velocity of the material, since the flow of air getting out of a subregion $A_i$ of the material is computed using an expression that is linear in the velocity $\dot{u}(x,t)$ of the material, i.e. $\mathcal{F}_i[\dot{u}]=\int_{A_i} -{\rm{div}} (\dot{u}) dx$.
Our analysis shows that the $\tau(t)$ frequency $f_o$ that maximizes the average airflow corresponds to the maximal velocity of the material and is equal to its fundamental frequency $f_1$.
Hence, the optimal frequency found by our analysis corresponds to the velocity resonance frequency of our fluid-structure interaction model.
\cb

This result suggests that the knowledge of the lung's characteristics could allow to optimize the therapy by computing the eigenfrequency of the material.
Actually, velocity resonance frequencies (i.e. eigenfrequencies) of the respiratory system have been estimated in the literature to about $6$ Hz for healthy adults \cite{mead_measurement_1956} and to about $18$Hz for infants lungs with respiratory distress syndrome \cite{lee_determination_1999}. 
Those estimations are close to the optimal frequency $f_o=6.5$ Hz obtained in our work by considering a characteristic length of $L=20$ cm for adult lungs. 
More particularly, if we assume approximately a characteristic length of $7$ cm for infants lungs, we obtain an optimal frequency of $f_o=18.5$ Hz.

Hence, our model gives for the first time a physical estimation of the optimal working range of HFCWO, which is in agreement with the frequencies usually applied to the patient during HFCWO maneuvers \cite{nicolini_effectiveness_2020}.\\ 

\noindent{\bf Influence of HFCWO on the mucus at the optimal frequency.}
The mucus stands on the wall of the airways as a thin layer of about $10 \ \mu$m~\cite{karamaoun_new_2018}.
Mucus is a viscoelastic fluid whose main property is to exhibit a yield stress that has to be overcome for the mucus to flow.
The order of magnitude of the yield stress $\sigma_0$ for an healthy mucus is typically $\sigma_0 \simeq 0.1$ Pa~\cite{lai_micro-_2009, mauroy_toward_2011, mauroy_toward_2015}.
During HFCWO manipulation, the mucus is submitted to two types of stresses: one arising from the air--mucus interactions and one from the oscillations of the airways walls.
These stresses add together and can either overcome directly the mucus yield stress and make it flow, or represent a buffer of stress, de facto reducing the quantity of stress to apply to overcome the mucus yield stress.\\

 \begin{figure*}[t!]
		\centering
		\centerline{\includegraphics[width=15cm]{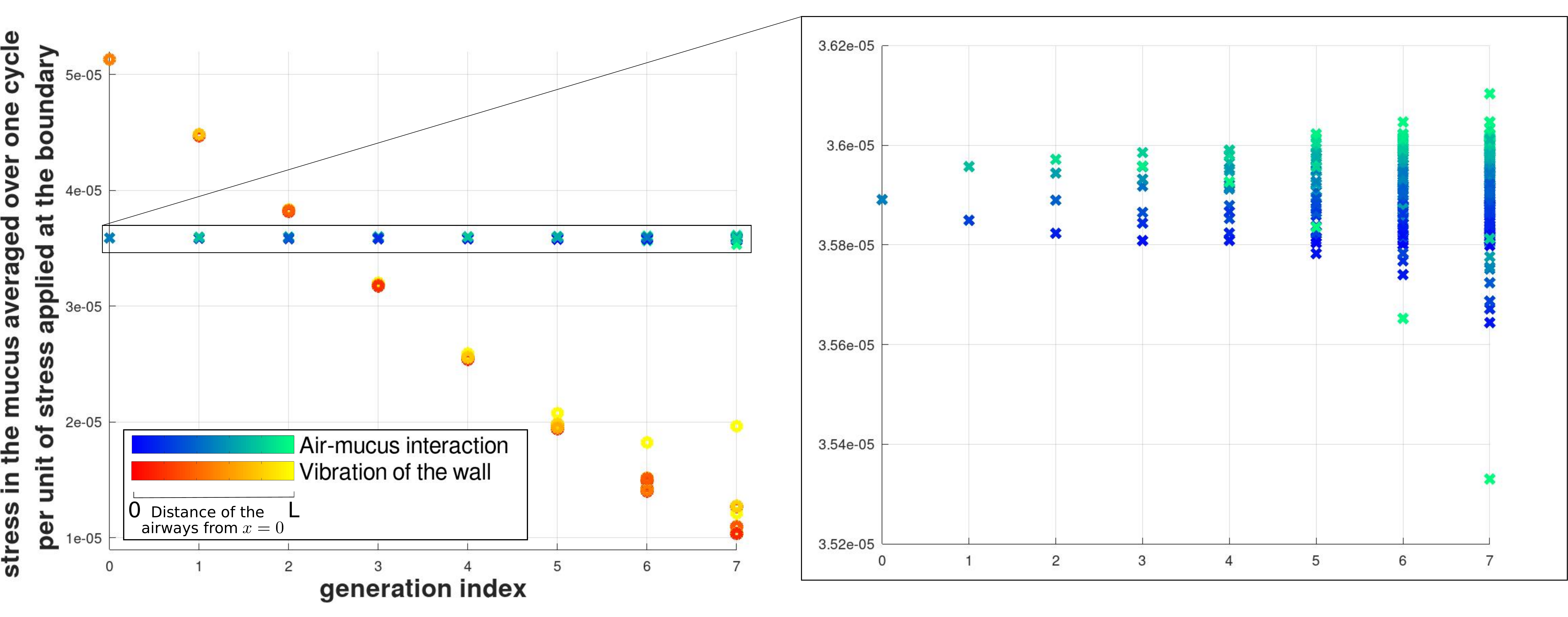}}
		\caption{Mean of the absolute wall shear stress (cold-colored crosses) and of the stress due to airways wall oscillations (hot-colored circles) {\cred per unit of applied pressure on the boundary, i.e. $\sigma_{a}/A$ and $\sigma_{\rm{w}}/A$.
		As our model is linear in $A$, the amplitude of the stress can be determined by multiplying the values plotted by $A$}.
		The mean is computed over the duration of one HFCWO cycle at the optimal frequency $f=6.5$ Hz, for all the bronchi of an eight generations tree. 
		The color reflects the location of the corresponding airway in $[0,L]$, i.e. the mean position of the $A_i$'s fed by the airway.
		For example, the root of the tree is feeding all the $A_i$ and its mean position is $L/2$; the two branches of the second generation are located at $L/4$ and $3L/4$, and so on.
		{\cred With the value $A=200$ Pa used in this work, the stress in the mucus represents between $10$ and $17\%$ of a typical yield stress of the healthy mucus \cite{lai_micro-_2009}.
		As expected, the modulation of the amplitude $A$ of the applied pressure allows to tune the amount of stress in the mucus.}
		}
		\label{F:6}
\end{figure*}

\noindent{\it Air--mucus interaction.}
The first stress is the one induced by the air--mucus interaction~\cite{mauroy_toward_2011, mauroy_toward_2015}.
As the mucus layer is in general thin relatively to the diameter of the airways, this stress can be approximated by the wall shear stress induced by the air flow in the airways~\cite{stephano_modeling_2019, stephano_wall_2021}.
As the airflows induced by HFCWO are small, we assume that the air fluid mechanics follows the Poiseuille's regime in the airways. 
Hence, the wall shear stress $\sigma_{\rm{a}}$ in an airway with radius $r$ and an airflow $\phi$ is~\cite{stephano_modeling_2019}
$$
\sigma_{\rm{a}} = \frac{\mu_a \phi}{\pi r^3}
$$
with $\mu_a$ the air viscosity, $\mu_a = 1.8 \ 10^{-5}$ Pa.s.
The wall shear stress in the tree is maximal when the air flow in the tree is maximal, typically for the optimal frequency uncovered previously.\\
	
\noindent{\it Airways wall oscillations.}
The tissue oscillations regularly compress and relax the airways, with the consequence of periodically affecting the geometry of the airways walls. 
The detailed derivation of the estimation of the stress occurring in the mucus is given in Appendix \ref{airwayRadius}.  
The airways deformations are small and the time evolution of their radii is determined based on the same model in \cite{mauroy_optimal_2008} that considers the airway walls as springs.
Then, we relate the radius of an airway $r(t)$ to the elastic properties of its wall and to the variations of the tissue pressures $p_t$ and of the air pressure in the airways $p_a$.
Finally, we assume that the Young's modulus $E_b$ of the walls of the airways is the same for all the airways and that $E_b = 6250$ Pa \cite{mauroy_optimal_2008}.
The pressure $p_a$ is taken as the mean air pressure in the airway and is computed using the pressure--flow relationships in the airways, see equation (\ref{eq1}).
The pressure $p_t$ is an estimation of the mechanical pressure surrounding the airway.
In our model, the airways have no spatial occupation, hence $p_t$ is estimated using the mean mechanical pressure over the region $\mathcal{Q}$ of the respiratory zone fed by the airway studied.
More precisely, if we consider all the paths from the terminal branches to the root of the tree, the set $\mathcal{Q}$ is the union of the $A_i$'s  that are coupled to a terminal branch whose associated path includes the airway studied.
Finally,
\begin{equation}
p_t(t) = \frac{\int_{\mathcal{Q}} \frac1m {\rm Tr}(\sigma(u)(t,x)) \ dx}{\int_{\mathcal{Q}} 1 \ dx}
\label{TrSigma}
\end{equation}
As our model is unidimensional in space, $m=1$ and $p_t(t) = \int_{\mathcal{Q}} (\lambda+2\mu) \frac{du}{dx} - p_{\rm{tree}}(\mathcal{F}[\dot{u}])\ dx / \int_{\mathcal{Q}} 1 \ dx$.

The way the radius evolves with time induces a tangential strain on the interface between the mucus and the airway wall, $\epsilon_{\theta}(r_0,\theta,z) = \frac{r(t)-r_0}{r_0}$.
Under the hypothesis that the pressure difference felt by the airway is small relatively to $E_b$, the resulting absolute stress in the mucus can then be estimated to, see Appendix \ref{stressMucus},
\begin{equation}
\sigma(t)_{\rm{w}} \simeq \frac12 \frac{r_0}{w_0} \frac{E_m}{E_b} \left|p_a(t) - p_t(t)\right|
\label{stressVib}
\end{equation}
with $w_0$ the thickness of the airway wall.
As suggested in \cite{preteux_modeling_1999, mauroy_optimal_2008}, the thickness can be approximated by $w_0 = \frac25 r_0$.\\

\noindent{\it Stress in the mucus in the optimal configuration.}
At the optimal frequency, we computed the absolute stresses averaged over one HFCWO cycle per unit of stress applied at the boundary to estimate its order of magnitude in the different airways of the tree, see Figure \ref{F:6}.  

The wall shear stress induced by the air--mucus interactions does not vary much along the generations and between the branches of the same generation.
Actually, the wall shear stress applied by the air on the mucus is directly related to the size of the airways and to the amount of airflow in the airway.
If the airflows were distributed equally in all the branches of a single generation, the shear stress should vary from one generation to the next with a factor $1/(2h_b^3) = 1$.
Consequently, in the hypothesis of a perfectly homogeneous distribution of airflows in the tree, the shear stress would be the same in all the generations of the tree, see \cite{stephano_modeling_2019, stephano_wall_2021}.
{\cred This behavior corresponds to Murray's law, originally expressed in the frame of cardiovascular fluid dynamics.
Murray's law determines the vessels geometry that minimizes the cost for blood transport and maintenance \cite{murray_physiological_1926}.}

However, our results indicate that there is a slight spread of the wall shear stresses that grows with the generation index.
This indicates that the difference of the airflows between the terminal branches are small relatively to the characteristic amplitude of the airflows in these branches.
Nevertheless, the airways that are closer to the boundary $x=0$, where the stress is applied, feel a stronger tissue pressure than the other airways.
Hence, they are submitted to larger stresses than the airways near $x = L$.

The stress due to the vibrations of the walls is larger than the stress induced by the air flow in the upper parts of the tree, but becomes smaller deeper in the tree.
This effect is related to the air pressure in the airways.
In the proximal part of the tree, the air pressure is small and the airways mechanics is mainly driven by the tissue pressure.
Hence, we can deduce from equation (\ref{stressVib}) that in the proximal airways, $\sigma(t) \simeq \frac54 \frac{E_m}{E_b} \left|p_t(t)\right|$.
Since the amplitude of $p_t$ is directly related to the applied sinusoidal stress of amplitude $A$, we can derive an estimation of the maximal possible mean stress over a cycle due to the wall vibrations of about $\sigma \simeq \frac54 \frac{E_m}{E_b} \frac{2 A}\pi = 0.025$ Pa when $A = 200$ Pa.
Although this quantity overestimates the stress found in our numerical simulations by a factor of about $2$, it remains of the same order of magnitude.
The shift was expected, as this approximation does not account for the real tissue pressure which depends on the wave propagation and on the damping by the tree.
Nevertheless, this approximation is a good way to get an estimation of the order of magnitude of the stress in the upper airways.
Deeper in the tree, the air pressure increases and compensates more strongly the tissue pressures around the airways.
As a consequence, the amplitudes of the oscillations of the airways walls decrease with the generations.

In HFCWO, the two stresses add together.
The idealised HFCWO technique used in our model, with $A = 200$ Pa, applies to the mucus a stress of about a hundredth of pascal, about ten percent of the yield stress necessary for an healthy mucus to move, whose yield stress is evaluated to be about $0.1$ Pa~\cite{lai_micro-_2009}. 
Adjusting the amplitude $A$ of our idealised HFCWO technique allows to reach higher stress in a proportional way, see Figure \ref{F:6}.\\

\section{Discussion}

We propose a mathematical and numerical model of the physics of HFCWO that highlights the empirical choices made for tuning HFCWO maneuvers.
	In our model, we account for the interaction between two core physical processes involved in the lung's biomechanics: the viscous dissipation of air in the airways and the mechanics of the deformation of the lung's tissues.
	The tree structure affects the displacement of the respiratory zone by applying damping pressures in the material. 
		Through the action of the air on the tree, a deformation affects the whole material very quickly --actually instantaneously in our model.
	The consequences predicted by our model for this dynamics is the existence of a range of frequencies for HFCWO that maximizes the airflow in the tree.
	This range of predicted optimal frequencies corresponds to the working frequencies empirically determined for HFCWO~\cite{nicolini_safety_2018, nicolini_effectiveness_2020}.
	In that range, the model suggests that the isolated action of a HFCWO therapy can submit an healthy mucus standing on the wall of the airways to about ten percent of the estimated yield stress that has to be overcome for the mucus to flow.
	In our model, this percentage can easily be tuned by adjusting the intensity of the applied pressure.
	
	Also, the physical analysis of our model suggests several interesting applications of the HFCWO technique.
	Actually, the optimal frequency can be determined by searching for the maximal airflow at mouth level.
	From that optimal frequency, it is possible to reach estimations of the hydrodynamic resistance and of the compliance of the patients lung, at least in the frame of our model.\\

\noindent{\bf Operational hydrodynamic resistance of the airway tree.}
We define the operational hydrodynamic resistance $R_{op}$ of the airway tree according to a distribution of the air flows at the terminal branches given by $\F$ and to the total air flow in the root (mouth air flow) given by $F_T = \phantom{} J^t \F$,
\begin{equation}
R_{op} = \int_0^T \phantom{} \F^t \mR \F dt \bigg/ \int_0^T F_T^2(t) dt
\end{equation}
The operational resistance reflects the resistance of the regions of the airway tree where there is an actual air flow.
Moreover, the influence of the regions is weighted according to the relative amount of airflows that they receive.
The regions where no airflow occurs are not accounted for in that resistance.
Hence, $R_{op}$ is in general an overestimation of the equivalent hydrodynamic resistance of the whole tree. 
In the case where the pressures at each terminal branches of the tree are similar, then the operational resistance is close to the equivalent hydrodynamic resistance $R_{eq}$ of the tree.

Coming back to our model of the lung, if we consider the balance of energy of the system of equations (\ref{eq2}), (\ref{eq3}), (\ref{eq3b}), (\ref{eq5}) over a cycle when the periodic regime is reached, then the energy dissipated during one cycle is equal to the amount of energy put in the system by the boundary $x=0$.
This balance is detailed in \ref{operApp} and can be summarized as
\begin{equation}
\int_0^T \tau(t) F_T(t) \ dt = \int_0^T \phantom{} \F^t \mR \F \ dt
\label{inv2m}
\end{equation}
The relationship (\ref{inv2m}) allows to estimate the operational resistance of a HFCWO maneuver if the applied signal $\tau(t)$ is known, if the total air flow $F_T(t)$ through the tree (i.e. the mouth airflow) is measured and if a periodic ventilation regime has been reached:
\begin{equation}
\label{opRes}
R_{op} = \int_0^T \tau(t) F_T(t) dt \bigg/ \int_0^T F_T^2(t) dt
\end{equation}
We showed earlier that in our idealised HFCWO maneuvers, the pressure jumps between the $A_i$ compartments are small relatively to the pressure itself, indicating that the pressures at the terminal branches are all similar in amplitude.
Hence, the operational resistance is a good approximation of the equivalent resistance of the tree in the case of the idealised HFCWO maneuvers.
Our numerical simulations confirms that during idealised HFCWO maneuvers, we have $R_{op} \simeq R_{eq}$. 

Hence, the operational resistance might have interesting applications for evaluating the actual resistance of the parts of the lung accessible to air flow, for evaluating the performance of a HFCWO maneuver and for estimating the equivalent hydrodynamic resistance of the lung using HFCWO.\\

\noindent{\bf Estimation of the compliance using the fundamental frequency.}
We showed that the optimal frequency in term of maximizing the mouth air flow is the fundamental frequency of the material.
This suggests that HFCWO could be used to estimate the compliance of the lung of a patient by searching for the device frequency that maximizes the air flow at mouth level.
Assuming this frequency is the fundamental frequency, we can derive from the formula $f_1 = \sqrt{\frac{\lm}{\rho}}\frac{1}{4 L}$ the elastic properties of the lung, represented here by $\lm$.
From $\lm$, we can estimate the lung compliance.
In the case of our unidimensional model, the compliance is related to the elastic parameters by $C \simeq V / \lm$ with $V = S_L L = L^3$ the volume of our model of the lung.
From the expression of the fundamental frequency, we can then deduce that
\begin{equation}
C \simeq \frac{V^{\frac13}}{16 \rho f_o^2}
\end{equation}
with $V$ the volume of the lung, $\rho$ its density and $f_o \simeq f_1$ the frequency that maximizes the air flow at mouth level.
This formula is derived from a unidimensional model and should be considered with care and/or be validated with clinical data.
However, this demonstrates that HFCWO might be a potential tool for estimating the lung's compliance based on the analysis of the air flows at mouth level.\\

	\noindent{\bf Model {\cred hypotheses}.}
	Our model predictions have to be interpreted in the limitations of its hypotheses.
	Actually, it is based on a set of simplification hypotheses for the geometry of the lung, the mechanics of the tissues and the air fluid dynamics. 
	
\cred	
The predictions of our model are based on averaged biological and mechanical parameters for a healthy individual and on an idealized self-similar bronchial tree geometry.
This allows to work with a tractable model and to identify the role of each biophysical phenomena on the dynamics of the system. 
However, the input parameters of the model exhibit inter-individual variations in the human population.
The bronchial tree geometry, the pulmonary resistance and the compliance are affected by the environment, the life-history, the age, the gender, etc. 
Accounting for this variability is possible with our model but is out of the scope of this study, which aims at analyzing the physics of the system.
Nevertheless, the influence on our model predictions of inter-individual variability could be analyzed in a future study using the same model.
Indeed, the first level model (generations $0$ to $n$) can be built using patient airways data, typically extracted from CT-scans \cite{tawhai_ct-based_2004}.
Then, patient compliance and hydrodynamic resistance could be used instead of the mean values used in this study.
To highlight the capacity of our model to run with patient data, a computation performed with Raabe et al. data \cite{raabe_tracheobronchial_1976} is presented in Appendix \ref{RaabeTree}.
\cb
	
\cred We assumed that the air flows in the airways according to a "corrected" Poiseuille regime, i.e. by adding a corrective factor for the Poiseuille hydrodynamic resistances of the airways in the first level model (generation $0$ to $n$).
This allows to account, in an approximated way, for the complex \cite{banko_oscillatory_2016, shang_detailed_2019, kim_CFD_2019}, sometimes chaotic \cite{tsuda_particle_2013, farghadan_topological_2019}, airflows in the upper airways (inertia, turbulence, influence of the geometry of the bifurcation, etc.).
The factor has been calibrated to get an equivalent resistance for the airway tree compatible with the physiology.
However, the flow patterns can be different depending on the airway size and orientation.
Hence, the use of a unique corrective factor for all the airways and for both rest ventilation and HFCWO cannot capture fully the complexity and variability of the fluid dynamics.
Hence, the local predictions of our model should be considered as qualitative only.
This might impact, for example, the predictions of the mucus stress induced by the wall oscillations, since the stress is computed using the predicted air pressures in the airways. 
Nevertheless, our approach allows to predict global behaviors and quantities that are coherent with the physiology, such as the optimal range of frequency for HFCWO.
This suggests that our model accounts, at least qualitatively, for the main biomechanical phenomena involved in HFCWO, including those arising from fluid dynamics.
\cb	

Another simplification was made concerning the physics of the system, for the sake of tractability.
Actually, the dimensionless parameter $\mathcal{E}$ is built from the equivalent hydrodynamic resistance of the tree. 
Hence, it captures only the mean influence of the dissipation of the energy by viscous effects in the tree.
Thus, some changes in the tree configuration can be missed as soon as the equivalent hydrodynamic resistance is not affected by the geometrical change.
To get a more fine description of the dynamics linked to the viscous dissipation in the tree, we can consider one dimensionless parameter per generation of the tree (symmetric branching) or per branch of the tree (asymmetric branching).
This improvement would allow to catch any influence of local changes in the tree, such as localized constrictions.
However, this would lead to a large number of dimensionless parameters, more than a hundred thousand for a $17^{\text{th}}$ generation tree, and would break the tractability of the model and its potential applicability to medicine.
	
		Nevertheless, our model is able to successfully mimic the rest ventilation and to capture the interactions between the tissue mechanics and the air flow in the airways. 
		The two dimensionless parameters $\mathcal{B}$ and $\mathcal{E}$ allow to highlight the relative influence of the elasticity and of the dissipation, depending on the physiological parameters and on how the idealised lung is ventilated.\\		
 

	\section{Conclusion}
	\label{S:6}
	
{\cred This work develops and analyzes a model of the lung that accounts for the main biophysical characteristics of the lung. 
The model is validated for rest ventilation by comparing its predictions for tidal volumes, mouth ariflows and alveolar pressures with data from the literature.}
The simulation of idealized HFCWO manipulations within this model of the lung brings estimations of the shear stress applied by the technique to the mucus.
We show that the stress that dominates in the upper part of the tree is the stress due to the vibration of the wall of the airways, while in the deep parts of the tree, the dominating stress is due to the air--mucus interactions.
We show that the frequencies ranging from $3$ Hz to $15$ Hz maximize the airflow inside the tree and consequently maximize the air--mucus interactions.
This range corresponds to the typical working frequencies empirically used during HFCWO.
Last but not least, in our model, the analysis of the mouth air flow during idealized HFCWO allows to estimate the hydrodynamic resistance and the compliance of our model of the lung. 
This suggests that HFCWO might be a powerful non invasive tool for helping the diagnosis of lung pathologies in the frame of personalized medicine.
	
Nevertheless, it is important to interpret our model predictions in the limits of our model hypotheses.
This work represents a first stepping stone toward the full understanding of the biomechanisms and the potential of HFCWO.
Further works will aim to reach more detailed prediction by improving the model realism, typically the geometry of the lung and the air fluid mechanics in the airways.


\section*{Acknowledgements}

This work has been funded by the Agence Nationale de la Recherche (VirtualChest, ANR-16-CE19-0014) and the Agence Nationale de la Recherche et de la Technologie (ANRT 2017/1493).

\section*{Conflict of interest disclosure}

M. Brunengo PhD thesis has been funded by RespInnovation SAS (Seillans, France) in the frame of the Association Nationale de la Recherche et de la Technologie (ANRT) that supports private/public research (CIFRE program).
B. R. Mitchell is a president at RespInnovation SAS (Seillans, France).

\section{Data availability statement}

The data that support the findings of this study are available from the corresponding author upon reasonable request. 

\onecolumngrid

\section*{References}

\bibliographystyle{abbrv}


\onecolumngrid

\newpage

\appendix

\begin{center}
\Large \bf Appendices
\end{center}

\section{Weak formulation and unidimensional case}
\label{weakApp}
\label{S:2:3}

The model of the lung developed in the previous section consists in the equation of the mechanics for the respiratory zone (\ref{eq2}), its boundary and initial conditions (\ref{eq3}), the viscoelastic stress--strain relationship for the model of the respiratory zone (\ref{eq5}), and the matrix pressures--flows relationship at the terminal branches of the bronchial tree model (\ref{eq1}).
The system is solved numerically using its weak formulation and finite elements.
The weak formulation and the finite elements method are convenient for dealing with the divergence of piecewise constant functions, such as the pressures $p_i$.\\

\noindent{\bf Weak formulation of the system of equations.} 
The equation that drives the mechanics of the tissue region is
\begin{equation}
\label{eq2App}
\left\{\begin{array}{ll}
\rho \frac{\partial^2 u}{\partial t^2} - \rm{div}(\sigma(u,\mathcal{F}[\dot{u}]))) = 0& \text{ $x \in \Omega$}\\
\mathbf{\sigma}(u,\mathcal{F}[\dot{u}])=\lambda {\rm Tr}(\epsilon(u))I+2\mu\epsilon(u)  - \ p_{\rm{tree}}(\mathcal{F}[\dot{u}])I& \text{ $x \in \Omega$}\\
\sigma(u,\mathcal{F}[\dot{u}])).n = \tau(x,t) & \text{ $x \in \Gamma_1$}\\
u(x,t) = 0 & \text{ $x \in \Gamma_2$} \\
u(x,0) = u_0(x) & \text{ for $x \in \Omega$}
\end{array}
\right.
\end{equation} 
The pressure $p_{\rm{tree}}$ is not everywhere differentiable, hence a relevant mathematical way to express the equation \eqref{eq2App} is by using the weak formulation.
For any proper smooth test function $w: \Omega \rightarrow \mathbb{R}^3$ which cancels on $\Gamma_2$, the weak formulation of \eqref{eq2App} is obtained by integrating on $\Omega$ the inner product of the equation with the test function $w$ and by applying the Stokes theorem.
The weak formulation of \eqref{eq2} is then
\begin{equation}
\label{eq6}
\left\{
\begin{array}{ll}
\int_{\Omega} \left(\rho \frac{\partial^2 u}{\partial t^2} w + \sigma_{\rm{e}}(u):\nabla w \right) dx - \int_{\Gamma_1} {\tau }.\ w  \ dS - \overset{N}{\underset{i=1}{\sum}} p_i(\mathcal{F}[\dot{u}]) \int_{A_i} {\rm{div}} \left( w \right) dx = 0 & \text{ on $\Omega$}\\
u(x,t=0) = u_0(x) & \text{ for $x \in \Omega$}\\
u=u_b  \ \text{ \bf  and } w=0 & \text{ on $\Gamma_2$} \\
 \left(p_i(\mathcal{F}[\dot{u}])\right)_{1\le i\le N}=-\left(\overset{N}{\underset{j=1}{\mathop{\mathlarger{\sum}}}} \mathcal{R}_{ij}\mathop{\mathlarger{\int}}_{A_j} {\rm{div}}(\frac{\partial u}{\partial t})dx \right)_{1\le i\le N} 
\end{array}
\right.
\end{equation}

\noindent{\bf Unidimensional case.}
In order to analyse the physics of the set of equations \eqref{eq6} in a tractable framework, we focus our study on unidimensional cases and limit the spacial dimension to the axis $x_1$.
The unidimensional geometry can be viewed in the three dimensional space as a cylinder that is the extrusion along the axis $x_1$ of a surface in the plane $(0,x_2,x_3)$. 
In terms of mechanics, we assume that the virtual displacement $w_1$ depends only on $x_1$ and that the displacements $w_2$ and $w_3$ are zero. 
We assume also that $\epsilon_{11}(u)$ is the single non zero term in the strain tensor.
For the sake of simplicity, we drop most of the index $1$ and use respectively $x$, $u$, $w$ and $\tau$ instead of $x_1$, $u_1$, $w_1$ and $\tau_1$. 

We assume that the extruded surface is a square with side length $L$ and surface area $S_L = L^2$.
Then, the rate of volume change can be rewritten
$\mathcal{F}_j[\du] = - \int_{A_j} \rm{div}(\frac{\partial u}{\partial t}) d x_1 d x_2 d x_3 = - S_L \int_{a_{j}}^{b_{j}} \frac{\partial}{\partial x}(\frac{\partial u}{\partial t}) dx$ where the projection of $A_j$ on the axis $x_1$ is the segment $[x_j, x_{j+1}]$. 
For the sake of simplification, we will now identify the set $A_j$ with its projection on the $x_1$ axis, i.e. $A_j = [x_j,x_{j+1}]$. 
Then, the tree pressure in the stress--strain relationship reformulates as
$$
\begin{aligned}
p_{\rm{tree}}(\mathcal{F}[\dot{u}]) = &\left(p_i(\mathcal{F}[\dot{u}])\right)_{1\le i\le N}\\
&= -S_L \left(\overset{N}{\underset{j=1}{\mathop{\mathlarger{\sum}}}} \mathcal{R}_{ij}\int_{x_j}^{x_{j+1}} \frac{\partial}{\partial x}(\frac{\partial u}{\partial t})dx \right)_{1\le i\le N} \\
&= -S_L \left(\overset{N}{\underset{j=1}{\mathop{\mathlarger{\sum}}}} \mathcal{R}_{ij}\left(\frac{\partial u}{\partial t}(x_{j+1},t)-\frac{\partial u}{\partial t}(x_j,t)\right) \right)_{1\le i\le N}
\end{aligned}
$$
where $\mR_{ij}$ is the $i,j$ component of the matrix $\mR$.

The definition \eqref{eq5} of the elastic stress tensor in 1D is $\sigma_{\rm{e}}(u)=(\lambda + 2\mu)\frac{\partial u}{\partial x}$.

\section{Dimensionless formulation of the equations and physical analysis}
\label{dimApp}

 The space, the time and the amplitude of the solution are adimensionalized as follows 
\begin{equation}
\label{eqeight}
\begin{array}{lll}
y = x/L\\
s = t/T\\
u(x,t) = \Upsilon\tilde{u}(y,s) = \Upsilon\tilde{u}(\frac{x}{L},\frac{t}{T})\\
p_i(\mathcal{F}[\du])=\mathcal{P} \tp_i(\tilde{\mathcal{F}}[\frac{\partial \tilde{u}}{\partial s}])   
\end{array}
\end{equation}
The quantities $L$ and $T$ represent respectively the characteristic length of the system and its characteristic time. 
The quantity $\Upsilon$ represents the characteristic displacement of the structure and $\mathcal{P}$ the characteristic pressure.
The space domain becomes $\tilde{\Omega}=[0,1]$.
$\tilde{\Omega}$ is decomposed into $N$ subsets $\tilde{A_i} = [\tx_i, \tx_{i+1}]$, which are the transformations by the adimensionalization of the corresponding $A_i$ in the original space.  
The rate of volume change is now 
$$\tmF[\frac{\partial  \tilde{u}}{\partial s}]=\left(-\int_{\tx_i}^{\tx_{i+1}}\frac{\partial}{\partial y}(\frac{\partial \tilde{u}}{\partial s})dy \right)_{1\le i \le N} = \left( \frac{\partial \tilde{u}}{\partial s}(\tx_i,s) - \frac{\partial \tilde{u}}{\partial s}(\tx_{i+1},s) \right)_{1\le i \le N}$$ 
We define the characteristic velocity $v$ to cross the whole system in a time $T$ as $v = L/T$.

The characteristic pressure $\mP$ is obtained by
$$
p_{\rm{tree}}(\mF[\pu]) = \mR \mF[\pu] = \mP \tp_{\rm{tree}}(\tmF[\ptu])  = \underbrace{R_{eq} \frac{S_L \Upsilon}{T}}_{\mP} \underbrace{\frac{\mR}{R_{eq}}\tmF[\ptu]}_{\tp_{\rm{tree}}(\tmF[\ptu])}
$$
where we recall that $R_{eq} = 1/( J^t \mR J)$ is the equivalent resistance of the tree, i.e. how it responds to an homogeneous distribution of pressures in its terminal branches. Hence, we can now define $\mP = R_{eq} \frac{S_L \Upsilon}{T}$. 
The quantity $\frac{S_L \Upsilon}{T}$ represents a characteristic air flow in the system.

The stress--strain relationship becomes,
$$
\tilde{\sigma}(\tu,\tmF[\ptu]) = \frac{\partial \tu}{\partial y} - \frac{\mathcal{P} L}{(\lambda + 2 \mu) \Upsilon} \ \tp_{\rm{tree}}(\tmF[\ptu]) = \frac{\partial \tu}{\partial y} - \frac{R_{eq} S_L L}{\lm T} \ \tp_{\rm{tree}}(\tmF[\ptu])
$$
and consequently, $\sigma(u,\mathcal{F}[\du]) = \frac{\Upsilon \lm}{L} \tilde{\sigma}(\tu,\tmF[\frac{\partial \tu}{\partial s}])$. We call the number $\mL_M = \frac{R_{eq} S_L v}{\lm}$ the Lung Mechanics number, it compares the characteristic pressure in the terminal branches $p_L = R_{eq} S_L v$ induced by the viscous dissipation of the air flow in the bronchial tree to the elastic response of the material, here represented by $\lm$.

At the boundary $y=0$, $\tsigma(\tu,\tmF[\frac{\partial \tu}{\partial s}]).n = \frac{A L}{\lm \Upsilon} \ttau(s)$, with $A$ the characteristic amplitude of pressure applied on the boundary and $\ttau(s)=\frac{\tau(Ts)}{A}$ the applied stress. 
In order to get a dimensionless stress at the boundary, we set the characteristic displacement to $\Upsilon = \frac{A L}{\lm}$. 
It is the result of the trade-off between the applied boundary stress and the elastic response of the material, scaled by the size of the object. 
Typically, the applied stress $\tau$ on the boundary is a sinusoidal signal with frequency $f$, $\tau(t) = A \sin(2 \pi f t)$, hence unless stated differently, $T = 1/f$ is the characteristic time of the system. 

Substituting these dimensionless quantities in the weak formulation of the system brings a new dimensionless weak formulation, for any smooth function $w$ such as $w(1) = 0$,
\begin{equation}
\label{eq9}
\int_{0}^1 \frac{\partial^2 \tu}{\partial s^2} w + \left( \mathcal{B} \frac{\partial \tu}{\partial y} - \mathcal{E} \tptree \right) \frac{\partial w}{\partial y}\ dy
- \mathcal{B} \ \underbrace{\int_{\tilde{\Gamma}_1} \ttau(s) w \ dy}_{= \ttau(s)w(0) \text{ (1D case) }}
=0
\end{equation}
with $\mathcal{B}$ the inverse of the Cauchy number of the system $\mathcal{B} =\lm / \rho v^2  $ that compares the elastic forces in the material with the inertial forces. 
The number $\mathcal{E} = \mL_M\mathcal{B}$ is actually the Euler number of the system since it can be rewritten in the form $\mathcal{E} = \frac{p_L}{\rho v^2}$. 
It compares the pressures forces induced by the viscous dissipation of the air flow in the bronchial tree with the inertial forces in the material.

\section{Numerical simulations, validation of the algorithm}
\label{numSim}

Numerical simulations using finite elements for the space variable $x$ are performed in {\it Octave} \cite{eaton_gnu_2019}. 
The time dynamics is computed numerically with the {\it Octave} function \rm{ode15s}.
The code is available in \cite{brunengo_code_2021}.

We determined analytical solutions of the unidimensional equations in the case of specific oscillating boundary conditions.
Analytical solutions are determined by decomposing the solution on each $A_i$.
We assume that their form on each $A_i$ is the product of a time-only-dependant function and of a space-only-dependant function.
The analytical solutions on $\Omega = \cup_{i=1}^N A_i$ is then obtained by assuming the continuity of the displacements and of the global stresses between two neighborhing $A_i$.

Our algorithm, which is able to deal with any boundary conditions, is then validated by comparing its predictions to these analytical solutions.

\section{Effect of the tree structure on the propagation of the wave deformation}	 
\label{S:3:2}
\label{S:3:2:1}

In the absence of the tree structure, i.e. with $\mathcal{E}=0$, the equation \eqref{eq9} is the linear elasticity equation for an isotropic and homogeneous material written in a dimensionless and uni-dimensional formulation. 
When the tree is present, it applies uniform pressures in the $A_i$, hence we expect that the tree structure will affect the displacements derivatives on the boundaries of the $A_i$.
Hence, we compare in this appendix the displacement of the material with or without the coupling with a tree structure using numerical simulations. 
This analysis allows to check the influence of the tree on the eigenfrequencies of the system, most particularly in term of resonance velocity.

The dimensionless domain is $\tilde{\Omega}=[0,1]$ and corresponds to a homogeneous material coupled with a three generations tree structure, as shown in Figure \ref{F:2}.\\ 

\begin{figure*}[t!]
		\centering
		\includegraphics[width=15cm]{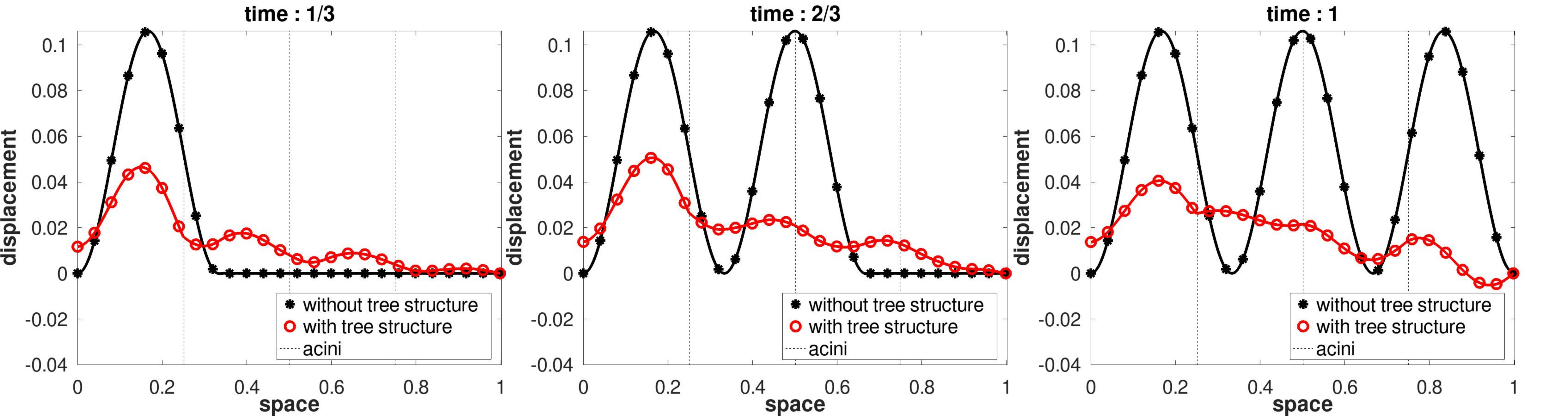}
		\caption{Propagation of wave deformation with (*-red) and without (o-black) the tree structure constraint at time $s=\frac{1}{3}$ (a), $s=\frac{2}{3}$ (b), $s=1$ (c), with $\mathcal{B}=1$ and $\mathcal{E}=1$. The dash vertical lines show the boundaries of the $(A_i)_{1\le i \le N}$.}
		\label{F:3}
\end{figure*} 

\noindent{\bf Boundary conditions.} 
The material is fixed (zero displacement) on one boundary ($y=1$) and stimulated by an oscillating pressure on the other ($y=0$). 
The pulsations $\omega_1 = \frac{\pi c}{2 L}$ and $\omega_2 = \frac{3 \pi c}{2 L}$ denote respectively the first and the second angular eigenfrequencies of the system without the tree, with $c = \sqrt{\lm / \rho}$, see \cite{rao_vibration_2007}. 
We set the Dirichlet condition $\tilde{u}(1,s)=0$ and the Neumann condition $\sigma(u).n (x,t) = \tau(t) = A \sin(\omega_2 t)$ rewritten in dimensionless formulation with $s = t/T$ and $T = 2 \pi / \omega_1$, $\tilde{\sigma}(\tilde{u})(y,s).n = \tilde{\tau}(s)=\tau(Ts)/A = \sin(\omega_2 T s)$ on $y=0 \;  \forall s \in \mathbb{R}^+ $.

We chose the pulsation of the boundary condition to be $\omega_2$ in order to be able to observe more easily the traveling wave.
As a consequence, a convenient characteristic time for the simulation is the time that the deformation wave takes to propagate from one boundary ($y=0$) to the other ($y=1$), which is $T=\frac{2\pi}{\omega_1}$.\\

\noindent{\bf  Initial conditions.} 
A zero initial condition is imposed on the displacements and velocities, the material is initially at rest.\\

\noindent{\bf Values of the physiological and physical parameters.}\\
The aspect ratio of the trachea is larger than that of the other airways.
Since all the airways sizes are computed from the size of the first generation, in order to compute satisfactory airway length, we have to use in our model a first generation airway that corresponds to a reduced trachea. 
Hence, the size of the first generation airway is assumed of length $l_0 = 6$ cm and of radius $r_0 = 1$ cm.
The resistance matrix $\mathcal{R}$ is then computed from these two values. 
In this section, we set the dimensionless parameters to  $\mathcal{B}=1$ and $\mathcal{E}=1$ ($\mathcal{E}=0$ for non-coupled case).\\

\noindent{\bf Wave propagation and dissipation.}
The black curves ({\bf *}-curves) in figure \ref{F:3} show the propagation of a wave in the absence of the tree, namely considering $\mathcal{E}=0$ . 
As the characteristic time is $T=\frac{2\pi}{\omega_1}$ with $\omega_1 = \frac{\pi c}{2 L}$ the fundamental angular frequency \cite{rao_vibration_2007}, the wave propagates through the material without any loss of energy and reaches the other boundary at time $s=1$.\\
The red curves ({\bf o}-curves) in figure \ref{F:3} show the propagation of the wave coupled to the tree structure ($\mathcal{E}=1$). 
The wave is damped by the air viscous dissipation occurring in the tree. 
Moreover, several areas of the domain are deformed before the arrival of the deformation wave. 
Actually, since we use Poiseuille's model for the air fluid mechanics in the tree, any change in pressures and airflows propagates instantaneously throughout the tree.
Hence, all the material is instantaneously affected by the change of the air properties in the $A_i$.

\section{Energy of the system}
\label{S:3:2:2}

In equation \eqref{eq9}, each term plays a part on the shape of the solution and is weighted and compared with the other terms by the value of its respective prefactor.
The prefactors are either $1$ or one of the two dimensionless parameters $\mathcal{B}$ or $\mathcal{E}$. 
The equation \eqref{eq9} can be decomposed according to the physical role of its different terms:
\begin{equation}
\int_0^1 \underbrace{\frac{\partial^2 \tu}{\partial s^2}}_{\text{acceleration}} w + \left( \smash[b]{\underbrace{\mathcal{B} \frac{\partial \tu}{\partial y}}_{\text{elasticity}}} - \smash[b]{\underbrace{\mathcal{E} \tptree}_{\text{damping}}} \right) \frac{\partial w}{\partial y}\ dy
- \underbrace{\mathcal{B} \  \ttau(s)}_{\text{boundary force}}w(0)
=0
\label{energy0}
\end{equation}
Taking $w$ in equation as the velocity of the material, i.e. $w = \frac{\partial \tilde{u}}{\partial s}$, we can determine the time variation of the energy of the system.
\begin{equation}
\frac{d}{ds} \underbrace{\left( \frac12 \int_0^1 \left(\frac{\partial \tilde{u}}{\partial s}(y,s)\right)^2 dy  
+ \mathcal{B} \frac12 \int_0^1 \left(\frac{\partial  \tilde{u}}{\partial y}(y,s)\right)^2 dy \right)}_{\text{Total energy of the system}}
= \underbrace{\mB \ttau(s) \frac{\partial \tilde{u}}{\partial s}(0,s)}_{\text{input power}} 
+ \underbrace{\mathcal{E} \ \sum_{i=1}^N \tp_i(\tilde{\mathcal{F}}[\frac{\partial  \tilde{u}}{\partial s}]) \tilde{\mathcal{F}}_i[\frac{\partial \tilde{u}}{\partial s}]}_{\text{viscous power dissipated}}
\label{energy}
\end{equation}
The total air flow through the tree is $J^t \tilde{\mathcal{F}}[\frac{\partial  \tilde{u}}{\partial s}] = \sum_{i=1}^N \tilde{\mathcal{F}}_i[\frac{\partial \tilde{u}}{\partial s}] = - \int_0^1 \frac{\partial^2 \tilde{u}}{\partial y \partial s}(y,s) dy = \frac{\partial \tilde{u}}{\partial s}(0,s) - \frac{\partial \tilde{u}}{\partial s}(1,s) = \frac{\partial\tilde{u}}{\partial s}(0,s)$.
Hence, the input power can be rewritten $\mB \ttau(s) \frac{\partial \tilde{u}}{\partial s}(0,s) = \mB \ttau(s) J^t \tilde{\mathcal{F}}[\frac{\partial  \tilde{u}}{\partial s}]$.

Since the pressure $\tp_i(\tilde{\mathcal{F}}[\frac{\partial  \tilde{u}}{\partial s}])$ in an $A_i$ is of the opposite sign than the corresponding flow $\mathcal{F}_i[\frac{\partial \tilde{u}}{\partial s}]$, then the term in $\mathcal{E}$ corresponds to a damping of the tissue, as expected.

\section{Model calibration and validation using rest ventilation}
\label{restApp}
\label{S:4:1}

We use our model to mimic the ventilation of the human lung. 
However, several parameters need to be adjusted in order for our model to give predictions compatible with the physiology.
The ventilation at rest in human is thoroughly studied in the literature, hence it is used to calibrate our model, see section \ref{S:4:1}. 
Once calibrated, our model is used to mimic HFCWO manipulation.

In this subsection, we mimic the pulmonary ventilation at rest. 
We consider the domain $\Omega=[0,L], L\in \mathbb{R}$ composed of an homogeneous material that mimics the lung's tissue. 
Here, we decompose the domain into $128$ subdomains $(A_i)_{i=0,\dots,127}$ which are fed by a tree of eight generations.
The deepest generations of the lung are mimicked using equivalent resistances added at each terminal branches of the eight generations tree. 

Our model of the bronchial tree is idealized and does not take into account the oesopharyngeal pathway, the detail of the geometry of the bifurcation and the inertial effects of the air flow~\cite{pedley_energy_1970, pedley_flow_1971}.
As a consequence, the hydrodynamic resistance of the tree and the resulting damping of the tissue deformation are underestimated if we base its computation on the geometry of our idealized tree only.
Hence, the hydrodynamic resistance of the idealized tree needs to be adjusted in order to get pressures and airflows compatible with the physiological values.
For that purpose, we introduced an ad-hoc corrective factor $c$ of $20$ for the hydrodynamic resistance of each branch of the tree of the first level of modelling.
{\cred The equivalent resistance $R_{eq}$ of the tree with the adjusted resistances is then equal to $1$ cmH$_2$O.L$^{-1}$.s.
This value is in accordance with the physiological data that estimates healthy adults hydrodynamic resistance to range from $0.5$ to $4$ cmH2O.L$^{-1}$.s \cite{maury_respiratory_2013}.
With this corrective factor, our model is then validated by ensuring that its predictions for tidal volume, alveolar air pressures and mouth airflows at rest are compatible with physiological data \cite{weibel_pathway_1984}.}\\

\noindent{\bf Parameters values.}
Since the lung's parenchyma is filled with $10\%$ of tissue and $90\%$ of air, the volumetric mass density $\rho$ of the material is set to $10\%$ of the volumetric density of water, i.e. $\rho = 100$kg/m$^3$ \cite{pozin_tree-parenchyma_2017, fuerst_personalized_2012}. 
This value for the density is probably not adapted to large volume variations of the lung, for which the air--tissue ratio could be significantly affected, typically during forced expiratory/inspiratory maneuver. 
However, for normal ventilation and especially for HFCWO conditions, we can reasonably assume that an air--tissue volume ratio of $90\%$ is a good approximation.

We use the same Young's modulus $E = 1256$ Pa and Poisson's ratio $\nu = 0.4$ as in \cite{pozin_tree-parenchyma_2017}.
From these data, we can compute the quantity $\lambda+2\mu$ used in our model using the equivalency between $(E, \nu)$ and $(\lambda, \mu)$ and the relationships $\lambda =\frac{E\nu}{(1-2\nu)(1+\nu)}$ and $\mu=\frac{E}{2(1+\nu)}$.

The resistance matrix $\Rm$ is built using physiological data for trachea radius and length: $r_0=1$ cm and $l_0=6$ cm.

We set the length $L$ to be compatible with the characteristic size of an adult lung, namely $L=20$ cm.

Finally, the amplitude of the boundary constraint $A$ is set to $200$ Pa. 
This value for $A$ allows our model to predict values for air flows and pressures in the airway tree that are fully compatible with the physiology.\\

\noindent{\bf  Initial conditions.}
The initial condition ($t=0$ s) corresponds to the material being still, i.e. no initial displacement and no initial velocity.\\

\begin{figure*}[t!]
		\includegraphics[height=5cm]{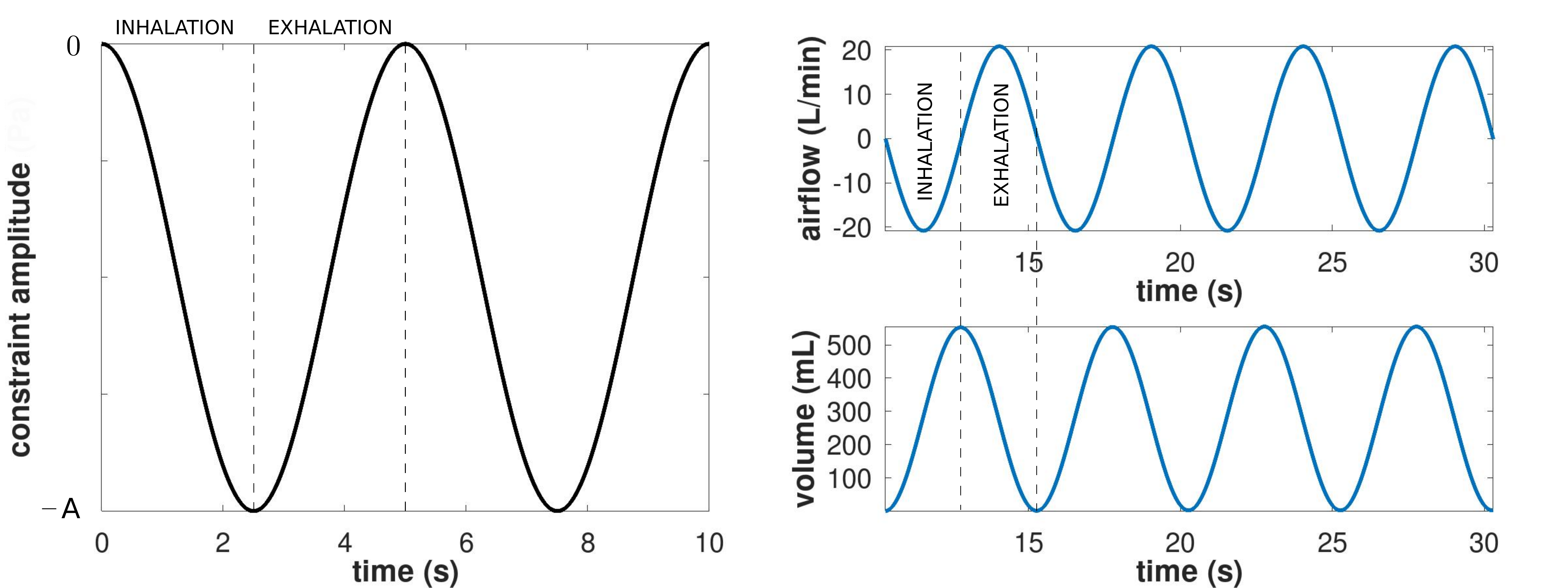}
		\caption{{\bf Left:} Pressure applied to mimic the action of the diaphragm (two respiratory cycles are shown). 
		The signal is sinusoidal in time with a period of $5$ seconds and an amplitude $A$. 
		{\bf Right:} Mouth airflow and tidal volume during four respiratory cycles of 5 seconds each with the adjusted hydrodynamic resistance. 
		The airflow and tidal volume data are displayed after they have reached a stationary state ($t\geq10$s).}
		\label{F:4}
\end{figure*}

\noindent{\bf Boundary conditions.}
The Neumann condition at $x=0$ is adjusted to mimic the pressure applied by the diaphragm to the lung during rest ventilation, $\sigma(u).n=\frac{A}{2}(cos(\omega t)-1), \text{ with } A\in \mathbb{R}$ and $\omega=\frac{2\pi}{5}$, or, in dimensionless formulation, at $y=0$, $\sigma(\tilde{u}).n=-\mathcal{B} (cos(2\pi s)-1)$. 
It mimics a negative pressure that moves back and forth the material, see Figure \ref{F:4} (left). 
The angular frequency $\omega$ is chosen so that the duration of a ventilation cycle is of $T = 5$ seconds (2.5 seconds inspiration and 2.5 seconds expiration) mimicking an idealized symmetric ventilation at rest~\cite{weibel_pathway_1984}. 
Moreover, we assume a zero displacement Dirichlet condition at $y=1$.

The numerical computations are performed with the dimensionless system of equations and the dimensional quantities are reconstructed from the dimensionless ones. 
Also, to go from computed 1D quantities to interpretable 3D quantities, we involve a surface of the material, denoted $S_L$ in section \ref{S:2:3}. 
We focus our analysis on the variations of the mouth airflow $\mF_{m}$ along time as this quantity is easily measurable in a clinical frame. 
The mouth airflow $\mF_{m}$ is computed as the sum of all the airflows in the terminal branches of the tree, 
\begin{equation}
\label{eq10}
    \mF_{m}\left[\frac{\partial u}{\partial t}\right]
= S_L\frac{\Upsilon }{T}\sum_{i=1}^N \tilde{\mF}_i\left[\frac{\partial \tilde{u}}{\partial s}\right]
\end{equation}

At rest, a human ventilates around $6$ to $8$ L/min~\cite{horwat_debit_1998} and a tidal volume of about $500$mL per respiratory cycle. 
With the adjusted resistance, our model predicts a tidal volume of around $553$mL.  
These results are shown in figure \ref{F:4} (right).
    
The values of dimensionless parameters associated to this case are $\mB = 16821$ and $\mE = 1043$. 
The Euler number $\mathcal{E}$, that represents the damping is about $16$ lower than the inverse Cauchy number $\mathcal{B}$ that represents the relative role of the elasticity.
This indicates that the system tends to dissipate slowly the elastic energy injected by the diaphragm.
Also, both $\mathcal{E}$ and $\mathcal{B}$ are larger than $1$, indicating that the acceleration plays a small role on the dynamics.

\section{Energy balance and operational resistance}
\label{operApp}

We consider the dimensionless energy conservation in our model, see equation (\ref{energy}) in Appendix \ref{S:3:2:2}.
We recall that the total air flow through the tree is $\tilde{F}_T(s) = J^t \tilde{\mathcal{F}}[\frac{\partial  \tilde{u}}{\partial s}] = \sum_{i=1}^N \tilde{\mathcal{F}}_i[\frac{\partial \tilde{u}}{\partial s}] = - \int_0^1 \frac{\partial^2 \tilde{u}}{\partial y \partial s}(y,s) dy = \frac{\partial \tilde{u}}{\partial s}(0,s) - \frac{\partial \tilde{u}}{\partial s}(1,s) = \frac{\partial\tilde{u}}{\partial s}(0,s)$.
Hence, the dimensionless energy balance is
\begin{equation}
\frac{d}{ds} \left( \frac12 \int_0^1 \left(\frac{\partial \tilde{u}}{\partial s}(y,s)\right)^2 dy  
+ \mathcal{B} \frac12 \int_0^1 \left(\frac{\partial  \tilde{u}}{\partial y}(y,s)\right)^2 dy \right)
= \mB \ttau(s) \tilde{F}_T(s)
- \mathcal{E} \ \tilde{F}^t \frac{\mR}{\mR_{eq}} \tilde{F}
\label{energy_3}
\end{equation}
since $ \tF^t \frac{\mR}{\mR_{eq}} \tF = - \sum_{i=1}^N  \tp_i(\tilde{\mathcal{F}}[\frac{\partial  \tilde{u}}{\partial s}]) \tilde{\mathcal{F}}_i[\frac{\partial \tilde{u}}{\partial s}]$.
Now, if we assume that the system is periodic with a period $1$, then integrating the previous equation over a cycle and going back to dimensional variables lead to
\begin{equation}
\int_0^1 \ttau(s) \tilde{F}_T(s) \ ds = \frac{\mL_M}{R_{eq}} \int_0^1 \phantom{} \tF^t \mR \tF \ ds
\ \rightarrow \
\int_0^T \tau(t) F_T(t) \ dt = \int_0^T \phantom{} \F^t \mR \F \ dt
\label{inv2}
\end{equation}
Hence, these relationships lead to define the operational hydrodynamic resistance of the airway tree 
\begin{equation}
R_{op} = \frac{\int_0^T \phantom{} \F^t \mR \F dt}{\int_0^T F_T^2(t) dt}
\end{equation}
When the pressures are all identical in the terminal branches of the tree, then we have $\phantom{} \F^t \mR \F dt = R_{eq} F_T^2$.
In that case, $R_{op} = R_{eq}$.
More generally, $R_{op}$ is the equivalent resistance of the parts of the tree where the air flows occur, weighted by the relative values of the airflows. 
Hence, the value of $R_{op}$ ranges from the value of the equivalent resistance of the most resistive path between the root of the tree and the terminal branches, and the value of the equivalent resistance of the tree.
In the quasi-fractal model with $n+1$ generations, $R_{op}$ ranges from $R_0+R_1+\dots+R_{n}$ (hydrodynamic resistance of a path from generation $0$ and $n$, all identical) and $R_{eq} = \sum_{i=0}^{n} \frac{R_i}{2^i}$.

\section{Extension of the definition of the dimensionless parameter $\mathcal{E}$}
\label{missing_information}

\subsection{Numerical simulation}
\label{A:C2}

\begin{figure*}[t!]
\begin{tabular}{cc}
\subfloat[Symmetrical 2 generations tree structure with $R_1=2 R_0$.]{\includegraphics[width=6cm]{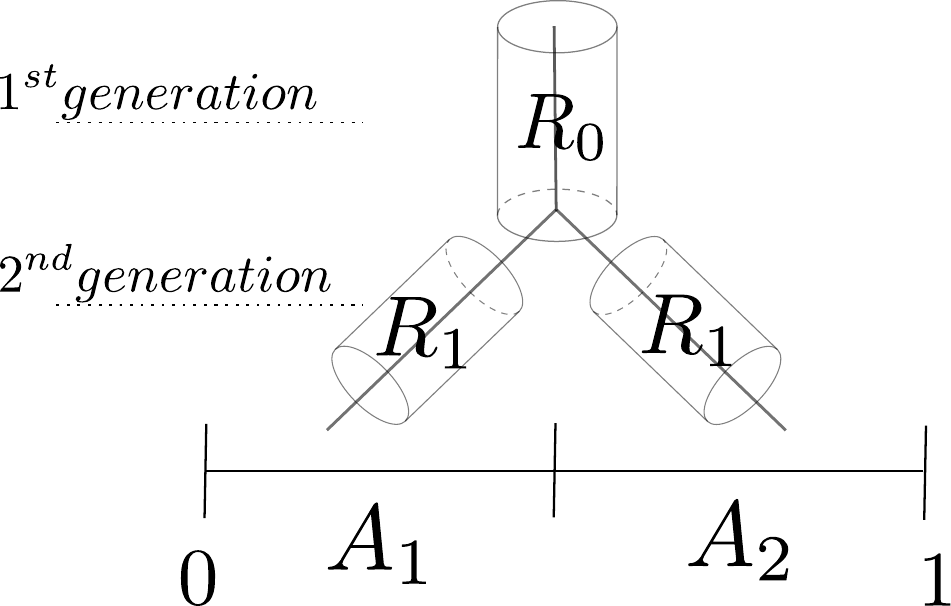}
\label{Symtreecomparison}
}
&
\subfloat[Non-symmetrical two generations tree structure with $R_{11}=R_1/100$ and $R_{12}=100 \, R_1$.]{\includegraphics[width=6cm]{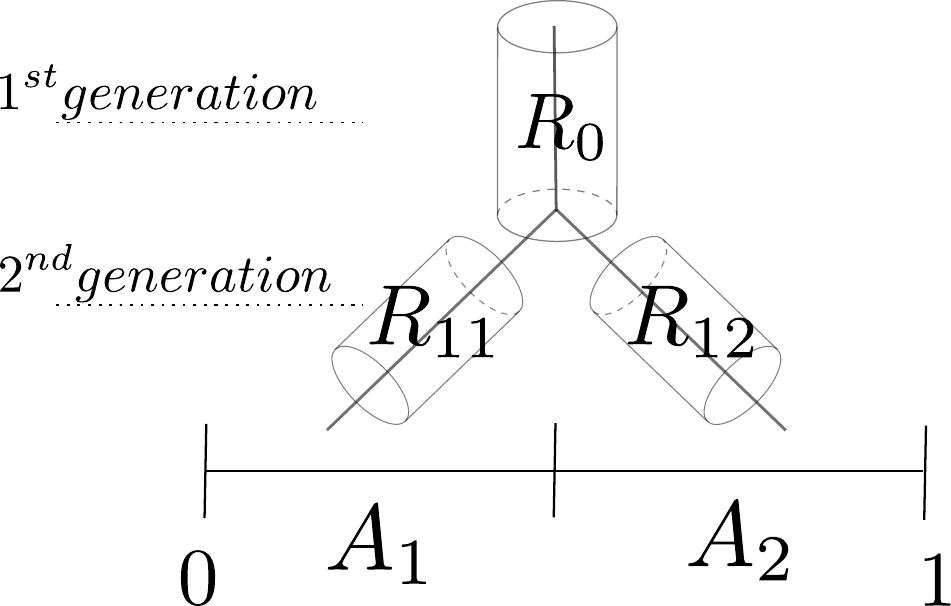} 
\label{Nonsymtreecomparison}
}
\\
&\\
\subfloat[Resistance matrix for a 2 generations tree structure with symmetric bifurcations.]{
$
\Rm = \left(
\begin{array}{cc}
\underbrace{R_0+R_1}_{3R_0} & R_0 \\
R_0 & \underbrace{R_0 + R_1}_{3R_0}
\end{array}
\right)
$}
&
\subfloat[Resistance matrix for a non-symmetrical 2 generations tree structure where $R_{11}=R_1/100$ and $R_{12}=100 \, R_1$.]{
$
\Rm = \left(
\begin{array}{cc}
\underbrace{R_0+R_{11}}_{(1.02)R_0} & R_0 \\
R_0 & \underbrace{R_0 + R_{12}}_{201R_0}
\end{array}
\right)
$
}
\end{tabular}
\caption{Two different tree structures with the same equivalent resistance and parameters $\mathcal{B}$ and $\mathcal{E}$, but with different resistance matrices. 
On the left is displayed a symmetrical tree structure (a) with the corresponding resistance matrix (c). 
On the right is displayed a non-symmetrical tree structure (b) with the corresponding resistance matrix (d).}
\label{T:C:6}
\end{figure*}
\begin{figure*}[t!]
\centering
\includegraphics[scale=0.13]{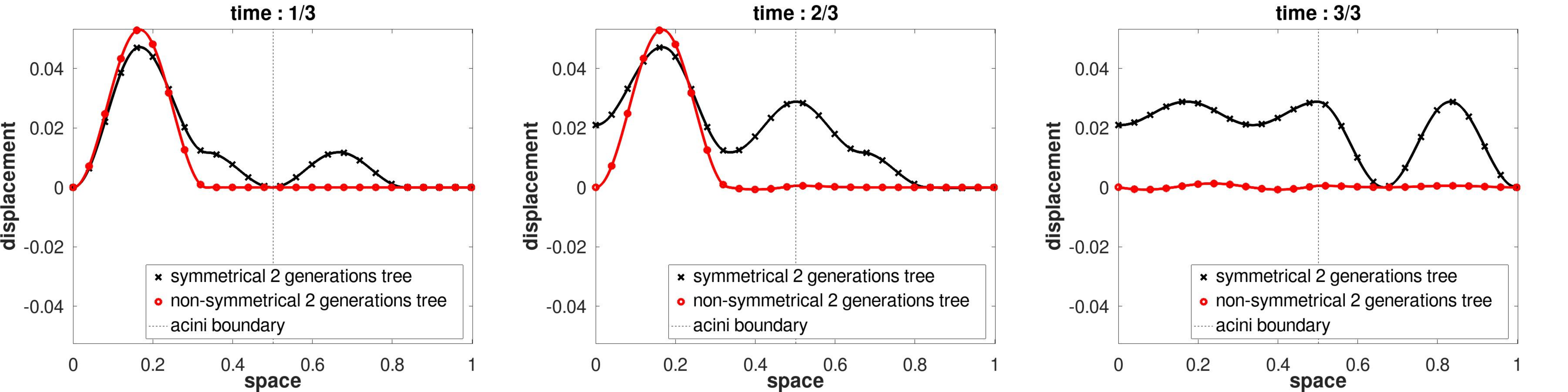}
\caption{Displacements for a two generations symmetrical (black cross) and a two generations non-symmetrical (red circle) tree with the same equivalent resistances and the same $\mB$ and $\mE$ at times $s=\frac{1}{3}$ (a), $s=\frac{1}{2}$ (b), $s=\frac{2}{3}$ (c).}
\label{D:7}
\end{figure*}
The dimensionless parameter $\mathcal{E}$ is not able to discriminate between all the trees, as two different trees can have the same hydrodynamic resistance but not the same branches.
To illustrate this phenomenon, we run two similar numerical simulations with the same $\mathcal{E}$, but with two different trees. 
We chose the resistance matrices $\Rm$ so that they are different but with the same equivalent resistance and consequently the same $\mathcal{B}$ and $\mathcal{E}$. 
The configuration chosen is based on a symmetrical and a non-symmetrical tree. 
We call $R_{11}$ and $R_{12}$ the hydrodynamic resistance of the two airways in the generation $2$.
The airway with resistance $R_{1i}$ feeds the set $A_i$ ($i=1,2$).
The non-symmetrical tree is related to the symmetric tree by a decrease by a factor 100 of the hydrodynamic resistance of one of the branch of generation $2$ --$R_{11}=R_1/100$-- and an increase by the same factor of the hydrodynamic resistance of the other branch --$R_{12}=100 \, R_1$--, see Figure \ref{T:C:6}. 
We keep $r_0=1$ cm and $l_0=6$ cm to define the hydrodynamic resistance $R_0$ of the root branch. 
The values of the dimensionless parameters are $\mathcal{B}=1$ and $ \mathcal{E}=1$.
We use the same boundary conditions as before and a zero initial condition on displacement and velocity. 

The Figure \ref{D:7} shows the material displacement for the two cases.
Since $A_2$ is connected to more resistive branch in the asymmetric case, the displacements in $A_2$ are lower in the non-symmetrical case than in the symmetrical case. 
This example shows that the local contribution of the resistance of each branch of the tree is not accounted for in the dimensionless formulation based on $\mE$ and $\mB$. 
This choice allows however to keep the physics of the system tractable and to connect more easily with clinical measures which often reflect global behaviors of the pulmonary system. 

\subsection{Alternative definition of the dimensionless parameter $\mathcal{E}$}
\label{A:C1}

The definitions of the dimensionless parameter $\mE = \frac{R_{eq} S_L}{\rho v}$ in Table \ref{T:1} is based on the equivalent hydrodynamic resistance of the tree $R_{eq}$ only.
This approach is not able to distinguish the dynamics induced by two different trees with the same hydrodynamic resistance.
Hence, we propose in this appendix alternative definitions of the dimensionless parameters.

In the case of a symmetrical bifurcating tree, one dimensionless variable $\mE_i$ can be defined for each generation $i$ of the tree.
In this case, all the branches belonging to the generation $i$ have the same hydrodynamic resistance $R_i$.
The dimensionless weak formulation is then
\begin{equation}
\label{eqD1}
\int_{0}^1 \frac{\partial^2 \tu}{\partial s^2} w + \left( \mathcal{B} \frac{\partial \tu}{\partial y} + \mathcal{E}^n  \; \tilde{F}_T \right) \frac{\partial w}{\partial y}\ dy
- \mathcal{B} \ \underbrace{\int_{\tilde{\Gamma}_1} \ttau(s) w \ dy}_{= \ttau(s)w(0) \text{ (1D case) }}
=0
\end{equation}
with $\mathcal{E}^n=\sum_{i=0}^n\mathcal{E}_i$, $\tilde{F}_T =\sum_{i=1}^N \tilde{\mathcal{F}}_i \left[\frac{\partial \tu}{\partial s} \right]$ and for $i=0,\dots,n$,
$$\mE_i = 2^{n-i} \frac{R_i}{R_{eq}}\mathcal{E}$$.

Similarly, in the case of a bifurcating tree with non symmetric bifurcations, all the branches of the tree can be different and one dimensionless number $\mE$ can be defined for each branch.
Hence, the number of dimensionless parameters $\mE$ would equal the number of branches in the tree, i.e. $2^{n+1}-1$ if the tree has $n+1$ generations.
The dimensionless number $\mE_b$ associated to a branch $b$ belonging to the generation $i$ and with a hydrodynamic resistance $R_b$ would then be 
$$\mE_b = 2^{n-i} \frac{R_b}{R_{eq}}\mE$$

It is important to adapt the number of dimensionless parameters to the problem in order to keep some tractability in the study.
	
\cred
\section{Fluid dynamics in an airway tree based on the physiological data from Raabe et al.}
\label{RaabeTree}

\begin{figure*}[t!]
		\centering		
		{\includegraphics[width = 19cm]{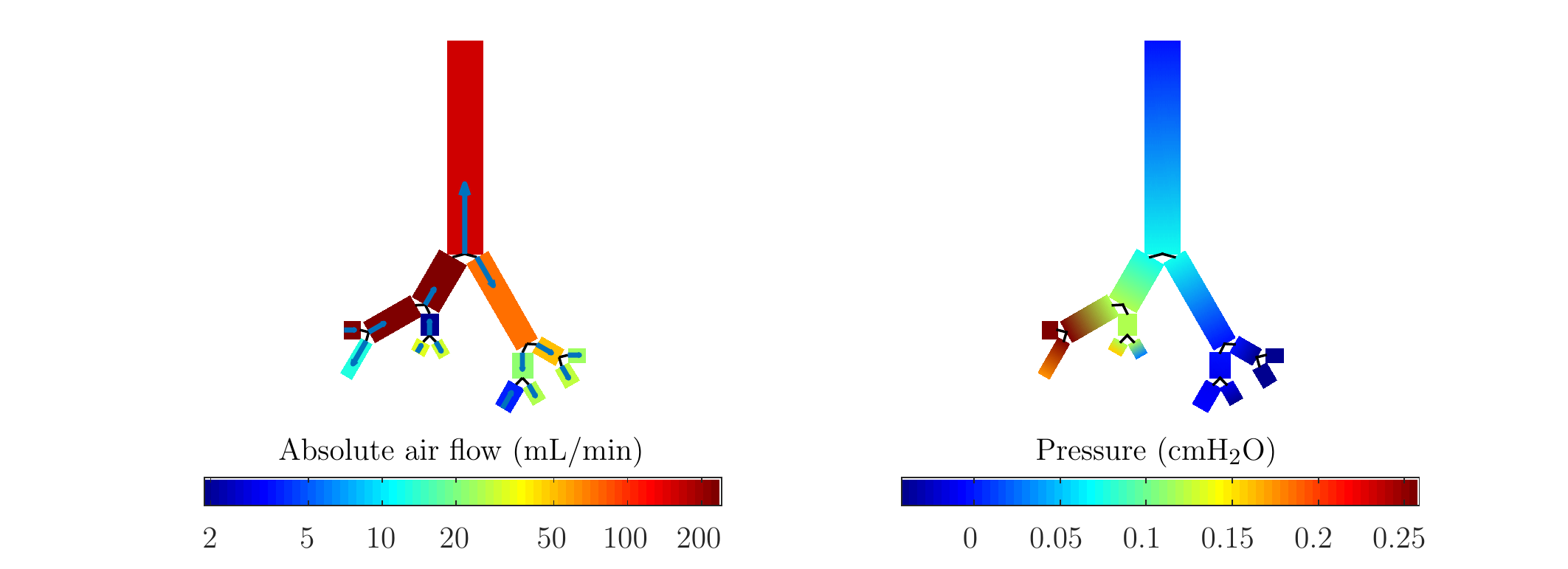}}
		\caption{\cred Air properties in the first level model of the bronchial tree with $4$ generations ($n=3$): mean airflows (up) and mean absolute pressures (down).
		The sizes of the airways are based on the data measured by Raabe et al. \cite{raabe_tracheobronchial_1976}.
		The blue arrows on the left plots represents the air flow orientation in the airways.
		The angles between the rectangles and the length of the arrows are chosen for visualization purpose only.
		The data are plotted for an idealized HFCWO frequency of $20$ Hz at the beginning of a HFCWO cycle. 
		This time corresponds to a maximal time derivative of the sinusoidal pressure applied at the position $x=0$ of the tissue (on the left).
		The rectangles represent the cylindrical airways to scale: their widths correspond to the airways diameters and their lengths to the airways lengths.
		The branches of the $4$-th generation account for the airways of the deeper generations, as schematized in Figure \ref{figTree}.
		The air circulates between different parts of the tree, getting away from the parts that are connected to the compressed regions of the tissue where the air pressure is higher (left part of the tree in the example plotted).
		The air is either expelled through the root of the tree (trachea) or sent to the parts of the tree connected to the regions of the tissue with less stress (right part of the tree in the example plotted).
		}
		\label{F:6Raabe}
\end{figure*}

		Our model is able to account for more realistic geometries based on measured data, typically using direct measurements \cite{raabe_tracheobronchial_1976} or airways reconstructions from CT-scans \cite{tawhai_ct-based_2004}.
		In this section, we propose an example using a tree of $4$ generations for the tree first level of modelling ($n=3$) built from the data measured by Raabe et al. \cite{raabe_tracheobronchial_1976}.
		The tree, the airflows and pressure in the airways of the $4$ first generations are represented in Figure \ref{F:6Raabe} at the beginning of a HFCWO cycle ($A=200$ Pa, $f = 20$ Hz).
		In Figure \ref{F:6Raabe}, the deformation wave propagates from the left ($x=0$) to the right ($x=L$).
		The fluid-structure interaction and the tree asymmetric bifurcations induces more complex airflows distributions than for the self-similar tree in Figure \ref{F:6}.
		The compression of the tissue in the left part of the tree induces high air pressure in the terminal branches feeding these regions of the tissue.
		A part of the airflow induced by the tissue deformation is going out of the tree through the generation $0$ that mimics the trachea.
		Another part of the airflow is going to terminal branches with lower pressures; as a result the tissue connected to these terminal branches tends to expand.
		Hence, the tissue deformation is the result of the propagation of the deformation wave in the tissue and of the distribution of the air flows in the tree.

\cb
	
\section{Estimation of the stress in the mucus layer induced by the airway walls oscillations}
\label{airwayRadius}
\label{stressMucus}

The airway walls are oscillating due to the oscillation of their transmural pressure. 
In order to estimate the stress applied on the mucus by the oscillations of the airway walls, we have first to determine the response of the airways radius to the changes in transmural pressure.
We assume that the airways wall behaves as a circular spring \cite{mauroy_optimal_2008}.
Then, we compute the stress in the mucus assuming that the mucus remains solid and behaves as a linear elastic material.
The hypotheses of linear elasticity are justified by the small amplitudes of the oscillations applied by HFCWO.

\subsection{Estimation of the evolution of the airways radii}

As HFCWO devices apply small deformations to the lung, we model the walls of the airways in the same way as in \cite{mauroy_optimal_2008}.
Hence, we assume that the airway wall reacts as a spring that remains circular.
We consider an airway with a rest radius $r_0$ and with a constant length $l_0$.
As in \cite{preteux_modeling_1999, mauroy_optimal_2008}, we assume the airway wall to have a thickness $w_0 = \frac25 r_0$, a Young's modulus $E_b = 6250$ Pa, a Poisson ratio $\nu_b = 0.5$ (incompressible material) and a density $\rho_b = 1000$ kg.m$^{-3}$.
We consider cylindrical coordinates $(r,\theta,z)$ adjusted to the cylindrical geometry of the airway: $r$ corresponds to the radial position, $\theta$ to the angular position and $z$ to the axial position.
The corresponding basis vectors are denoted $e_r$, $e_{\theta}$ and $e_z$.
Assuming that the cylindrical airway has a radius $r$, we consider a part of its wall with an angular width of $d\theta$ located at the angular position $\theta$.
Applying the Newton's second law to that segment leads to
\begin{equation}
\underbrace{r \, d\theta \, w_0 \, l_0 \, \rho_b}_{\text{mass}} \underbrace{\frac{d^2 r}{dt^2} \bfer(\theta)}_{\text{acceleration}}  = \underbrace{t(r) \, l_0 \, \bfeth(\theta) - t(r) \, l_0 \, \bfeth(\theta + d\theta)}_{\text{elastic force}} + \underbrace{r \, d\theta \, l_0 \left( p_a(t) - p_t(t) \right)\bfer(\theta)}_{\text{pressures forces}}
\label{PFD}
\end{equation}
where:
\begin{itemize}
\item 
The function $r \rightarrow t(r)$ is the lineic tension due to the elongation of the wall, $t(r) = - \frac{E_b}{1-\nu_b^2} w_0 \frac{r-r_0}{r_0}$, see more details in \cite{mauroy_optimal_2008}.

\item 
The pressure in the tissue $p_t(t)$ is computed using the trace of the stress tensor $\sigma(u)$ in the respiratory zone, see equation (\ref{eq5}),
$$
p_t(t) = \frac{\int_{\mathcal{Q}} \frac1n {\rm Tr}(\sigma(u)(t,x)) \ dx}{\int_{\mathcal{Q}} 1 \ dx}
$$
with $n$ the spatial dimension and $Q$ the set defined as the union of the $A_i$ fed by the airway studied.
For example, the tissue pressure in the first generation airway is the mean of the tissue pressures computed on all the $A_i$ as this airway is feeding all the tissue, i.e. $Q = \Omega$.
With $n=1$, we can rewrite the pressure in the tissue as 
$$
p_t(t) = \frac{\int_{\mathcal{Q}} (\lambda+2\mu) \frac{du}{dx} - p_{\rm{tree}}(\mathcal{F}[\dot{u}])\ dx}{\int_{\mathcal{Q}} 1 \ dx}
$$

\item
The air pressure in the airway $p_a(t)$ results from the air fluid dynamics in the tree.
It is approximated by the mean air pressure in the airway, which is computed using the linear relationships between the air flows and pressures in the tree, see equation (\ref{eq1}).
More precisely, we define the set $I$ of the indexes of the airways that are on the path starting from the root of the tree and ending at the airway studied.
For $i \in I$, we denote $R_i$ is the hydrodynamic resistance of the airway with index $i$ and $\phi_i$ the airflow in that same airway; the quantity $R_i \phi_i$ is the pressure drop in the airway $i$.
Finally, denoting $R_b$ the hydrodynamic resistance of the airway studied and $\phi_b$ the airflow in that same airway, we have
$$
p_a(t) = - \left( \sum_{i \in I} R_i \phi_i(t) \right) + \frac{R_b}{2} \phi_b(t)
$$
where the first term computes the pressure at the end of the airway studied and the second term is a correction to get the pressure in the middle of that airway.\\
\end{itemize}

Using the relationship $\bfeth(\theta) - \bfeth(\theta + d\theta) = \bfer(\theta) d\theta$, projecting the equation (\ref{PFD}) on $\bfer$ and simplifying, we obtain
\begin{equation}
w_0 \rho_b \frac{d^2 r}{dt^2} = -\frac{w_0}{r_0} \frac{E_b}{1-\nu_b^2} \frac{r-r_0}{r} + \left( p_a(t) - p_t(t) \right)
\label{PFD2}
\end{equation}

Then, rewriting the equation (\ref{PFD2}) in a dimensionless form allows to compare the different influences of acceleration, elasticity and pressures:
$$
\underbrace{\frac{\rho_b r_0^2 (1-\nu_b^2)}{E_b T^2}}_{N} \frac{d^2 \tilde{r}}{ds^2}  = 
- 1 \times \frac1{\tilde{r}}\left( \tilde{r} - 1 \right)  
+ \underbrace{\frac{r_0}{w_0} \frac{P (1-\nu_b^2)}{E_b}}_{M} \left(\tilde{p}_a(s) - \tilde{p}_t(s)\right)
$$
using $s = t/T$, $\tilde{r}(s) = r(sT)/r_0$, $\tilde{p}_*(s) = p_*(sT)/P$ with $*=a$ or $t$.
$T$ is the characteristic time of the oscillations, i.e. their period; at the optimal configuration (see figure \ref{F:5}), $T = 0.05$ s. $P$ is the order of magnitude of the pressure, typically the pressure applied on the boundary, reflected by the variable $A$, hence we chose $P = A = 200$ Pa.
Finally, we can estimate the dimensionless numbers $N = \frac{\rho_b r_0^2 (1-\nu_b^2)}{E_b T^2}$ and $M = \frac{r_0}{w_0} \frac{P (1-\nu_b^2)}{E_b}$ at the optimal configuration,
$$
N \leq N_{|r_0 = 1 \rm{cm}} = 5.1 \ 10^{-4} \ \text{ and } M = 6.0 \ 10^{-2}
$$ 
Consequently, the acceleration is small relatively to the elastic term ($N << 1$) and we can assume at first approximation a static equilibrium between the elastic forces and the pressure forces.
Notice that $N$ decreases when the generation index increases since the radii of the airways are decreasing with the generation index. 
Hence, the approximation $N << 1$ is better for the small airways.
The number $M$ is also quite small relatively to $1$, indicating that the displacements due to the pressures are also small, in agreement with the linear elasticity approximation.

Solving the static equation leads to $\tilde{r}(s) = (1 - M (\tilde{p}_a(s) - \tilde{p}_t(s))^{-1}$.
Considering $M<<1$, we can go further in the approximation, and $\tilde{r}(s) \simeq 1 + M (\tilde{p}_a(s) - \tilde{p}_t(s))$.
Equivalently, using dimensional variables and replacing $\nu_b$ with $1/2$, we can finally reach an expression for $r(t)$:
$$
r(t) = \frac{r_0}{1-\frac{3}{4} \frac{r_0}{w_0} \frac{p_a(t) - p_t(t)}{E_b}} \simeq r_0 \left( 1 + \frac{3}{4} \frac{r_0}{w_0} \frac{p_a(t) - p_t(t)}{E_b}\right)
$$

The determination of the evolution of the radius relatively to that of the transmural pressure $p_a(t) - p_t(t)$ allows to compute in the next section the resulting stress in the mucus.
 
\subsection{Estimation of the stress in the mucus layer}

The way the radius evolves with time induces a tangential strain on the interface between the mucus and the airway wall, $\epsilon_{\theta}(r^0,\theta,z) = \frac{r(t)-r_0}{r_0}$. 
This tangential strain propagates into the mucus at a characteristic velocity $c = \sqrt{E_m/\rho_m}$ where $E_m$ is the Young's modulus of the mucus and $\rho_m$ its density.
In an healthy mucus layer, $E_m \simeq 1$ Pa and $\rho_m \simeq 1000$ kg.m$^{-3}$~\cite{lai_micro-_2009} and we can estimate that $c \simeq 3$ cm.s$^{-1}$.
Since the typical thickness of the mucus layer is about $10 \ \mu$m~\cite{karamaoun_new_2018}, the wave propagates through the depth of the mucus in less than $0.5$ ms.
Hence, the strain on the mucus wall represents well the strain inside the mucus layer at the time scale of HFCWO.
At the position $(r, \theta, z)$ the strain in the mucus layer is then
$
\epsilon_{\theta}(r,\theta,z) = \frac{r(t)-r_0}{r_0}
$.
The mucus is an incompressible material, hence the trace of the strain operator is zero and $\epsilon_{\theta} = - \epsilon_r$ since we assume $\epsilon_z = 0$.
Finally, based on these hypotheses and on the linear elasticity in cylindrical coordinates states that
$$
\begin{array}{l}
\epsilon_r = (1+\nu_m)\left( (1 - \nu_m) \sigma_r - \nu_m \sigma_{\theta} \right) / E_m\\
\epsilon_{\theta} = (1+\nu_m)\left( (1 - \nu_m) \sigma_{\theta} - \nu \sigma_r \right) / E_m
\end{array}
$$
Then, using $\epsilon_r = - \epsilon_{\theta}$, we have $\sigma_* = \frac{E_m}{1+\nu_m} \epsilon_*$ with $*=r$ or $\theta$. 
Making the mucus Poisson's ratio $\nu_m$ going to $0.5$ since the mucus is incompressible, the norm of the stress in the thin layer of mucus on the wall of the airway can be estimated with
$$
\sigma(t) = \frac23 E_m \left|\frac{r(t)-r_0}{r_0}\right| \simeq \frac12 \frac{r_0}{w_0} \frac{E_m}{E_b} \left| p_a(t) - p_t(t) \right| 
$$ 

The stress induced in the mucus by airways wall oscillations is then compared to the yield stress of the mucus, see main text.

\end{document}